\documentclass[lettersize,journal]{IEEEtran}
\usepackage{algorithm}
\usepackage{algorithmicx}

\usepackage{algpseudocode}

\usepackage{CJK}

\usepackage{pifont}

\usepackage{amsmath}
\usepackage{amssymb}
\usepackage{psfrag}
\usepackage{graphicx}
\usepackage{epstopdf}
\usepackage{multirow}
\usepackage{stfloats}
\usepackage{cite}
\usepackage{subcaption}
\usepackage{color}
\usepackage[normalem]{ulem}
\usepackage{arydshln}
\usepackage{enumerate}
\usepackage{bbm}
\usepackage{verbatim}
 
\newtheorem{theorem}{\bf{Theorem}}[section]

\newtheorem{remark}{\bf{Remark}}
\newcommand{\bm}[1]{\mbox{\boldmath{$#1$}}}

\begin{document}

\title{Learning-Enabled Elastic Network Topology for Distributed ISAC Service Provisioning}
\author{
Jie Chen, \IEEEmembership{Member, IEEE}, and Xianbin Wang, \IEEEmembership{Fellow, IEEE}
\thanks{
This manuscript was submitted to \textit{IEEE Transactions on Cognitive Communications and Networking} on December 23, 2025; revised on April 13, 2026; and accepted on April 28, 2026.
This work was supported in part by the Natural Sciences and Engineering Research Council of Canada (NSERC) Discovery Grants Program under Grants RGPIN-2024-05720 and RGPIN-2026-05399, and in part by the Canada Research Chairs Program under Grant CRC-2023-00336. (Corresponding author: Xianbin Wang.)
 
J. Chen was with the  Department of Electrical and Computer Engineering, Western University, London, ON N6A 5B9, Canada. He is now with the Department of Electrical and Computer Engineering,
University of New Brunswick, Fredericton, NB E3B 5A3, Canada (e-mail: jie.chen@unb.ca).

X. Wang is with the  Department of Electrical and Computer Engineering, Western University, London, ON N6A 5B9, Canada (e-mail: xianbin.wang@uwo.ca).
}
}
 \maketitle
\begin{abstract}
Conventional mobile networks, including both localized cell-centric networks (CCNs) and cooperative cell-free networks (CFNs), are built upon rigid network topologies.
However, neither architecture is well-suited to flexibly support distributed integrated sensing and communication (ISAC) services, due to the increasing difficulty of aligning spatiotemporally distributed heterogeneous service demands with available radio resources.
In this paper, we propose an elastic network topology (ENT) for distributed ISAC service provisioning, in which multiple coexisting localized CCNs can operate independently within their local boundaries or be dynamically grouped into CFNs with
expanded boundaries for federated network cooperation.
This topology elastically orchestrates the boundaries of localized CCNs and federated CFNs to balance signaling overhead and distributed resource utilization, thereby enabling efficient ISAC service provisioning.
A two-phase operation protocol is then developed.
In Phase-I, each CCN autonomously partitions its resources into dedicated and shared segments and classifies ISAC services as either local or federated.
In Phase-II, each CCN employs its dedicated resources for local-type services, while the aggregated CFN consolidates shared resources from its constituent CCNs to deliver federated-type services cooperatively.
Furthermore, we design a utility-to-signaling ratio (USR) to quantify the tradeoff between service utility and the operational signaling overhead.
Consequently, a USR maximization problem is formulated by jointly optimizing the network topology (i.e., service classification and CCN grouping) and the allocation of dedicated and shared resources.
However, this problem is challenging due to its distributed optimization nature and the absence of complete channel knowledge.
To address this problem efficiently, we propose a multi-agent deep reinforcement learning (MADRL) framework with centralized training and decentralized execution. Simulation results validate the effectiveness of the proposed system design and algorithm development.
\end{abstract}


\IEEEpeerreviewmaketitle

\section{Introduction}

\subsection{Background and Motivation}
With the rapid convergence of wireless technologies and vertical applications such as industrial automation and intelligent transportation, future wireless networks are expected to simultaneously support both communication-centric and beyond-communication services at massive scales \cite{liu2022survey}.
Consequently, integrated sensing and communication (ISAC) has emerged as a key enabler of this vision by integrating sensing and communication functionalities within a unified infrastructure that leverages shared waveforms, hardware, spectrum, and energy resources \cite{zhang2021enabling,wei2023integrated,jia2025integrated}.
This integration enables efficient and user-specific operation, enhancing the network's ability to accommodate the stringent and often conflicting requirements of heterogeneous service provisioning, especially under constrained energy and spectral resources.

Currently, ISAC service provisioning is realized under two representative network architectures: cell-centric networks (CCNs) and cell-free networks (CFNs).
These two directions follow distinct design philosophies, leading to contrasting trade-offs in coordination complexity, signaling overhead, and service provisioning performance.
\begin{itemize}
\item The CCN-based local ISAC service provisioning adopts a decentralized architecture  \cite{li2024maximizing}, in which each cell independently manages local sensing and communication tasks using its own channel state information (CSI) and dedicated resources, without inter-cell coordination. This architecture significantly reduces signaling overhead from inter-cell CSI exchange, synchronization, and control operations.
However, such single-cell-based resource utilization under rigid cell boundaries suffers from strong inter-cell interference owing to the lack of cooperation, as well as inherent limitations in service coverage, user capacity, transmission reliability, and resource efficiency resulting from limited spatial diversity \cite{shan2024resource}. Hence, it faces significant challenges in adapting to rapidly changing wireless environments for dynamic service provisioning.

\item The CFN-based cooperative ISAC service provisioning enables centralized coordination and coherent signal processing \cite{chen2025radiation}, thereby mitigating inter-cell interference and enhancing spatial diversity, which in turn significantly improves the performance of sensing and communication tasks.
Nevertheless, full cooperation in CFNs requires CSI of all links among the associated transceivers \cite{chen2025knowledge}, resulting in substantial signaling overhead for CSI acquisition. Moreover, the coupled optimization of sensing- and communication-related variables leads to a high-dimensional decision space and increased optimization complexity, especially in large-scale CFNs. 
\end{itemize}

The limitations of rigid CCN and CFN topologies for spatiotemporally dynamic ISAC service provisioning motivate the development of an elastic network topology (ENT), which can 
orchestrate operations across localized CCN and federated CFN regimes by leveraging the strengths of both topologies to efficiently adapt to rapidly changing service requirements and resource availability in distributed ISAC deployments.

\subsection{Related Works}

Most existing studies on CCN-based ISAC service provisioning focus on intra-cell resource allocation for single base station scenarios, aiming to achieve various performance trade-offs between sensing and communication services.
These resource-sharing studies can be broadly divided into two categories.
The first type balances sensing and communication trade-offs within a single transmission frame without considering temporal channel correlation.
Specifically, communication performance is typically characterized by metrics such as capacity \cite{ren2023fundamental,xu2024anti}, outage probability \cite{zhang2023semi}, and bit error rate \cite{wu2024low}, while sensing performance is evaluated using detection probability \cite{an2023fundamental}, the Cram\'er--Rao bound (CRB) \cite{dong2022sensing,li2025integrated}, sensing mutual information rate \cite{xie2025sensing}, and beam pattern matching error \cite{hua2023optimal}.
In addition to these metrics, a performance measure called value-of-service was recently proposed in \cite{li2024maximizing,chen2025beyond} to capture the significance of individual sensing or communication service provisioning events, which can be applied to design fair resource allocation under competing demands and constrained resources.
The second type extends the design to multiple transmission frames and exploits temporal channel correlation to improve resource efficiency.
For example, the impact of channel aging on ISAC systems was investigated in \cite{chen2023impact,chen2024learning}, and the findings were further leveraged to guide resource allocation by extending the CSI estimation interval, thereby reducing signaling overhead while maintaining joint sensing and communication performance requirements.
In addition, extended Kalman filtering \cite{liu2020radar} and deep learning (DL)-based methods \cite{liu2022learning,zhang2024predictive} have been proposed to predict beamforming designs, thereby achieving the tradeoff between the Bayesian CRB and communication rates.

On the other hand, research on CFN-based ISAC service provisioning has been relatively limited, but has shown increasing activity in recent studies.
For example, transmit beamforming was studied in \cite{sun2025interference} for interference management to balance the communication rate and sensing signal-to-interference-plus-noise ratio (SINR), and in \cite{mao2024communication} for optimizing the communication–sensing region using a Lagrangian dual method.
A drift-adaptive slicing-based resource management scheme was further proposed in \cite{hu2025drift}, which jointly considers long-term and short-term timescale resource planning to maximize sensing coverage under communication rate constraints.
In addition, joint active access point (AP) activation with beamforming coordination was investigated in \cite{chen2025radiation} to minimize the radiation footprint of transmitters for autonomous interference control with reduced operational overhead, and in \cite{tung2025joint} to improve energy efficiency under rate and CRB constraints.
Finally, the number of antennas and the power allocation were optimized in \cite{meng2025network} to enable a sensing-communication trade-off at the network level.

However, the above designs are based either on an independent CCN or on a fully cooperative CFN and thus cannot effectively adapt to the envisioned ENT-based distributed ISAC systems proposed in this paper. Such scenarios demand dynamic resource allocation based on the elastic cooperation boundary between CCNs and CFNs to cope with rapidly changing service requirements and resource availability.

\subsection{Main Contributions}
Motivated by the above limitations, we develop an ENT-based network architecture that enables dynamic orchestration of network topology, resource allocation, and signaling overhead to efficiently support spatiotemporally varying ISAC service demands. The main contributions are summarized as follows:
\begin{itemize}
\item  
To the best of our knowledge, this is the first study to adopt a dynamic topology for ISAC service provisioning. Specifically, we propose an ENT for distributed ISAC service provisioning, in which multiple coexisting CCNs can operate independently within their local boundaries or be dynamically grouped into CFNs with expanded boundaries for federated network cooperation.
The proposed architecture flexibly orchestrates the topology between localized CCNs and federated CFNs to balance signaling overhead and radio resource utilization, thereby supporting highly efficient and adaptive service provisioning.

\item
We develop a two-phase operation protocol tailored for ENT-based distributed ISAC service provisioning.
During Phase I, each CCN autonomously classifies services into local or federated types and partitions its resources into dedicated and shared portions.
During Phase II, each CCN utilizes its dedicated resources for local service provisioning, while the aggregated CFN consolidates the shared resources of its constituent CCNs to cooperatively deliver federated services.

\item
We further design a utility-to-signaling ratio (USR) performance metric, defined as the ratio of the joint sensing and communication utility to the signaling overhead. This metric effectively captures the tradeoff between sensing/communication performance and signaling cost.
We then formulate a USR maximization problem that jointly optimizes the network topology (i.e., service classification and CCN grouping) and the allocation of dedicated and shared resources, including bandwidth, beamforming vectors, and transmit power.

\item The formulated problem is challenging due to its distributed optimization nature and the lack of complete channel state information. To efficiently address this issue, we develop a multi-agent deep reinforcement learning (MADRL) framework based on proximal policy optimization under the centralized training and decentralized execution (CTDE) paradigm. Simulation results verify the effectiveness of the proposed architecture and algorithm.
\end{itemize}

\textit{Organization}: Section II introduces the system model, and Section III derives the system performance metrics.
Section IV formulates the optimization problem, while Section V develops the corresponding solutions.
Finally, Section VI provides the simulation results, and Section VII concludes the paper.

\textit{Notation:} For a score vector $\mathbf z=[z_1,\ldots,z_L]^{\mathsf T}\in\mathbb{R}^{L}$ and a temperature $\tau>0$, ${\rm softmax}_{\tau}(\mathbf z)$ denotes the temperature-scaled softmax vector whose $\ell$-th entry is $\exp(z_\ell/\tau)\big/\sum_{\ell'=1}^{L}\exp(z_{\ell'}/\tau)$, for $\ell=1,\ldots,L$. For a probability vector $\boldsymbol{\pi}=[\pi_1,\ldots,\pi_L]^{\mathsf T}$ satisfying $\pi_\ell\ge 0$ and $\sum_{\ell=1}^{L}\pi_\ell=1$, the notation $x\sim{\rm Cat}(\boldsymbol{\pi})$ means that $x\in\{0,\ldots,L-1\}$ with $\Pr(x=\ell)=\pi_{\ell+1}$ for $\ell=0,\ldots,L-1$. Moreover, letting $u_\ell\sim{\rm Uniform}(0,1)$ and $g_\ell=-\log(-\log u_\ell)$ for $\ell=1,\ldots,L$, ${\rm Concrete}_{\tau}(\mathbf z)$ denotes the Concrete random vector whose $\ell$-th entry is $\exp((z_\ell+g_\ell)/\tau)\big/\sum_{\ell'=1}^{L}\exp((z_{\ell'}+g_{\ell'})/\tau)$, for $\ell=1,\ldots,L$.

\section{System Model}\label{sec:SystemModel}
\begin{figure}
\centering
\includegraphics[width=0.495\textwidth]{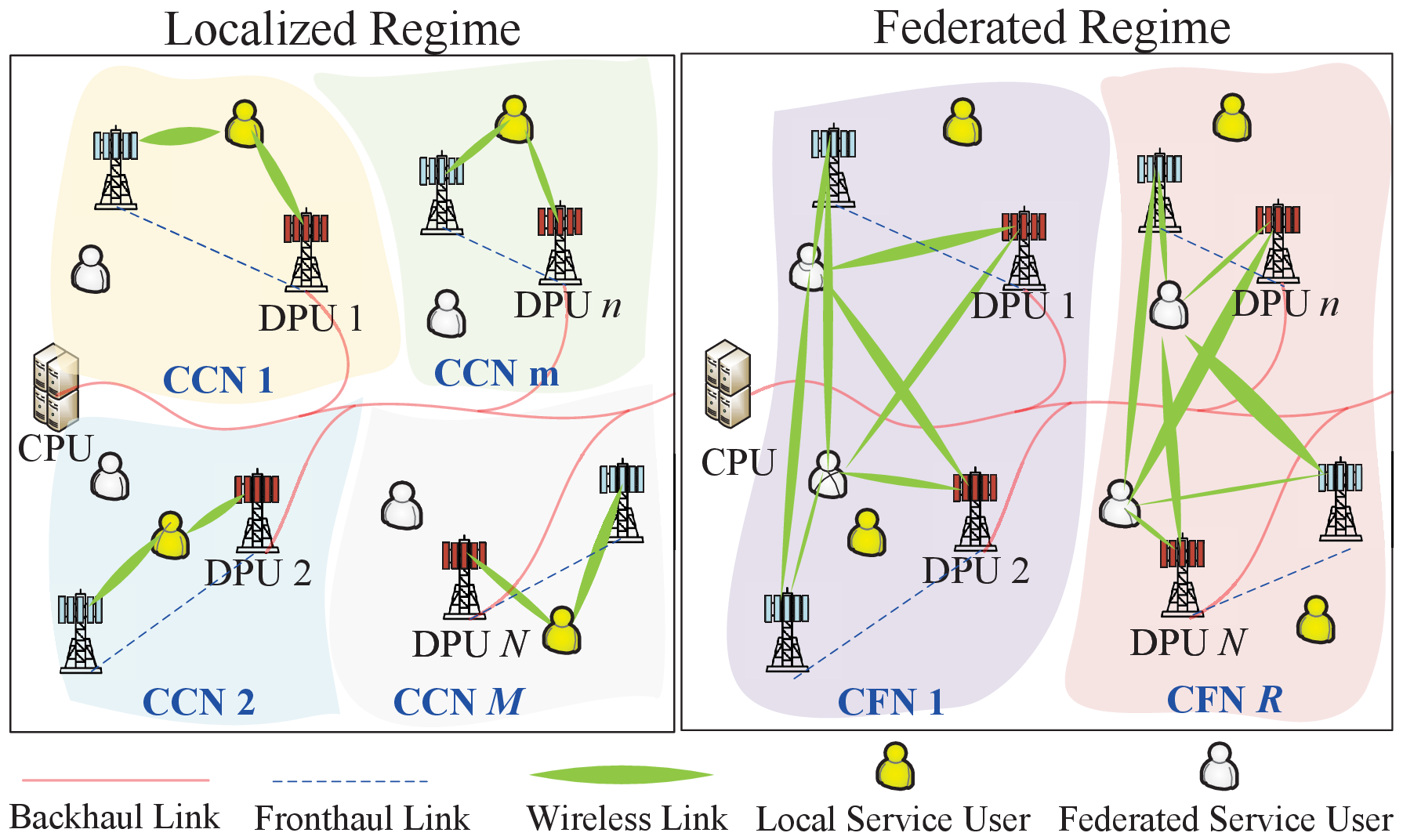}   
\caption{Illustration of the ENT-enabled distributed ISAC service provisioning. The network elastically adjusts its operational boundaries between  CCN-based localized regimes with lower signaling overhead and  CFN-based federated regimes with enhanced performance at the cost of higher coordination overhead.}\label{fig1SystemModel}
\end{figure}

As shown in Fig. \ref{fig1SystemModel}, we consider a distributed ISAC network consisting of $M$ CCNs, each comprising $A$ APs, $K$ communication users, and $Q$ radar targets.
Without loss of generality,   the transceiver nodes in CCN $m$, $1\le m\le M$, are represented by the tuple $\left\{ {\mathbb K}_m, {\mathbb Q}_m, {\mathbb A}_m \right\}$,  i.e.,
   
\begin{itemize}
\item ${\mathbb A}_m=\{a\mid (m-1)A+1\le a\le mA,\; a\in\mathbb Z\}$ denotes the index set of APs in the $m$-th CCN;
\item ${\mathbb K}_m=\{k\mid (m-1)K+1\le k\le mK,\; k\in\mathbb Z\}$ denotes the index set of communication users in the $m$-th CCN;
\item ${\mathbb Q}_m=\{q\mid (m-1)Q+1\le q\le mQ,\; q\in\mathbb Z\}$ denotes the index set of radar targets in the $m$-th CCN.
\end{itemize}
Each CCN includes one host AP equipped with a digital processing unit (DPU), while the remaining APs are non-host APs connected to the host AP via fronthaul links to support intra-cell local operations within the same CCN.
Additionally, all DPUs are connected to a central processing unit (CPU) via backhaul links to facilitate inter-cell federated cooperation.
We further assume that each AP operates in full-duplex mode and is equipped with $N_{\rm tx}$ antennas, while each user is equipped with a single antenna.
This system is operated over $N_{\rm T}$ time frames in time division duplex (TDD) mode, ensuring channel reciprocity within each frame \cite{abedi2024low}.
The total bandwidth is equally divided into $M$ subbands, with each subband containing $B$ subcarriers, where each subcarrier has a bandwidth of $\Delta_f$.

\subsection{Channel Model}

\subsubsection{Communication Channel}
Let the positions of communication user~$k$  and AP~$a$ be  ${\bf{x}}_{k}^{\rm{C}}= \left[ {x_{k}^{{\rm{C}},\rm{x}},x_{k}^{{\rm{C}}, \rm{y}}} \right]^{\mathsf T}$ and ${\bf{x}}_{a}^{\rm{A}} = \left[ {x_{a}^{{\rm{A}},\rm{x}},x_{a}^{{\rm{A}},\rm{y}}} \right]^{\mathsf T}$, respectively, where the superscripts ``$\rm C$'' and ``$\rm A$'' indicate that the terms are associated with the communication users and APs, respectively, and the superscripts ``$\rm x$'' and ``$\rm y$'' indicate the components along the x-axis and y-axis, respectively.
Then, the channel between communication user~$k$ and AP~$a$ on subband~$m$ in frame~$n$ is denoted by ${\bf{h}}_{akm}^{\rm{C}}[n]\in{\mathbb C}^{N_{\rm tx}}$. It is assumed to evolve across successive frames following a first-order Gaussian–Markov process \cite{chen2024learning}:
\begin{align} \label{eqchannelN1}
{\bf{h}}_{akm}^{\rm{C}}[n] = {\rho _{ak}}{\bf{h}}_{akm}^{\rm{C}}[n - 1] + \sqrt {1 - \rho _{ak}^2} {\bm{\varepsilon }}_{akm}^{\rm{C}}[n],
\end{align}
where ${\rho _{ak}}\in\left[0,1\right]$ denotes the correlation coefficient of the channel fading between adjacent frames and $
{ { \bm\varepsilon }}_{akm}^{\rm{C}}\left[ n \right] \sim {\cal C}{\cal N}\left( {0, {\lambda _{ak}^{\rm C}}{{\bf{I}}_{{N_{\rm tx}}}}} \right)$ denotes the evolving noise, which is independent of the channel. 
Here, $\lambda_{ak}^{\rm C}=10^{-{\rm PL}({\bf{x}}_k^{\rm C},{\bf{x}}_a^{\rm A})/10}$ denotes the large-scale fading coefficient, where ${\rm PL}(\cdot,\cdot)$ is the path-loss function in dB.
 We initialize  ${\bf{h}}_{akm}^{\rm{C}}[0]$ as
\begin{align}
\!\! {\bf{h}}_{akm}^{\rm{C}}[0]\!=\! \sqrt {\frac{{\lambda _{ak}^{\rm{C}}}}{{\bar \kappa \! + \!1}}} \left( \!{\sqrt {\bar \kappa {N_{{\rm{tx}}}}} \tilde h_{akm}^{{\rm{LoS}}}[0]{\bf{v}}(\theta _{ak}^{\rm{C}}) \!+ \!{\bf{\tilde h}}_{akm}^{{\rm{NLoS}}}[0]} \!\right),\label{eqchannel1}
\end{align}
where $\bar \kappa$ is the Rician factor. Besides,  ${\tilde h}_{akm}^{\rm{LoS}}\left[ 0 \right] \sim {\cal C}{\cal N}\left( {0,1} \right)$  and ${\bf{\tilde h}}_{akm}^{\rm{NLoS}}\left[ 0 \right] \sim {\cal C}{\cal N}\left( {0,{{\bf{I}}_{{N_{\rm tx}}}}} \right)$ represent the Line-of-Sight (LoS) and Non-Line-of-Sight (NLoS) components, respectively.
Moreover,  ${\bf{v}}(\theta _{ak}^{\rm{C}})$ is the steering vector 
with angle  $
\theta_{ak}^{\rm C}
=
\arctan\!\left(
\frac{x_a^{{\rm A},{\rm y}}-x_k^{{\rm C},{\rm y}}}
{x_a^{{\rm A},{\rm x}}-x_k^{{\rm C},{\rm x}}}
\right)$, i.e.,
$
{\bf{v}}\left( \theta  \right) = \frac{1}{\sqrt{N_{\rm tx}}}{\left[ {1,{e^{ - {\rm{j}}\pi \sin \theta }}, \cdots ,{e^{ - {\rm{j}}\pi \left( {{N_{{\rm{tx}}}} - 1} \right)\sin \theta }}} \right]^{\rm{T}}} 
$.

\begin{figure*}
 \centering
 \includegraphics[width=0.9\textwidth]{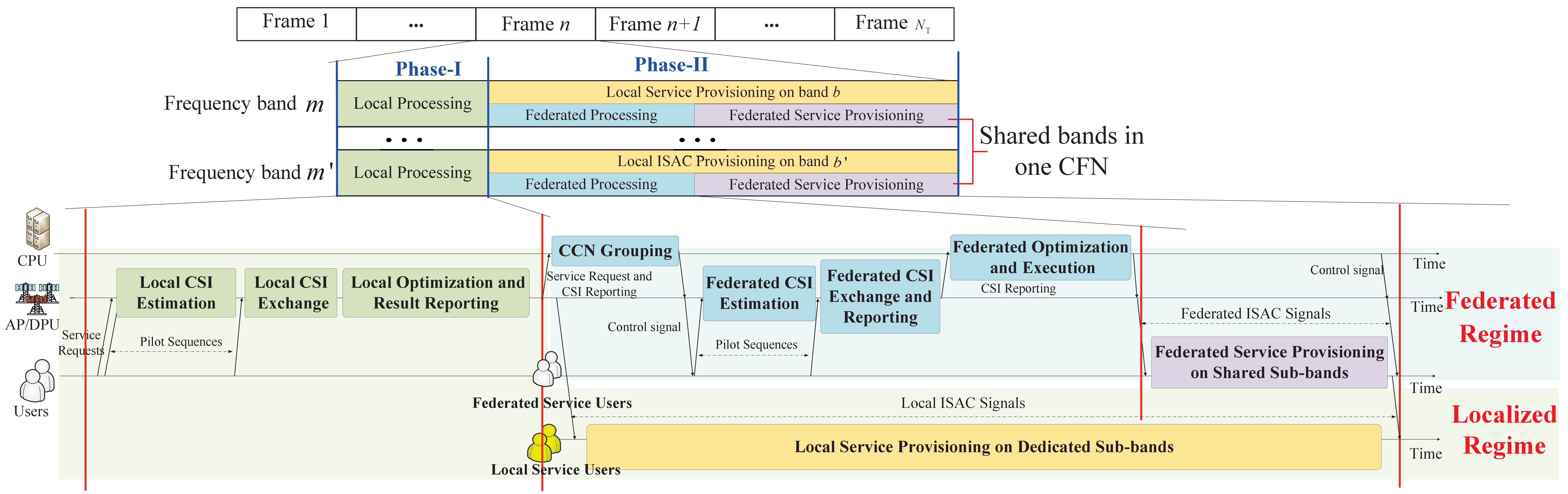}
 \caption{Transmission frame structure for the ENT-based distributed ISAC service provisioning system.}\label{fig2Protocol} 
\end{figure*}

\subsubsection{Target Mobility Model}

We assume that targets are located separately in a Cartesian coordinate system and we denote the position and velocity of radar target $q$ in frame $n$ by 
${\bf{x}}_{q}^{\rm R}[n]
=
\left[
x_{q}^{{\rm R},{\rm x}}[n],
x_{q}^{{\rm R},{\rm y}}[n],
v_{q}^{{\rm R},{\rm x}}[n],
v_{q}^{{\rm R},{\rm y}}[n]
\right]^{\mathsf T}$, where the superscript ``$\rm R$'' indicates that the terms are associated with radar targets.
Then, the motion of target $q$ is modeled by:
 \begin{align}
 {\bf{x}}_q^{\rm{R}}\left[ n \right] = {\bf{Gx}}_q^{\rm{R}}\left[ n -1\right] + {\bm{\varepsilon }}_q^{\rm{R}}\left[ n \right] \in{\mathbb R}^{4},\label{eq5}
 \end{align}
 where ${\bf G}$ is the mobility transition matrix, and ${\bm{\varepsilon }}_q^{\rm{R}}\left[ n \right]$ is Gaussian noise with mean zero and covariance ${\bf{E}}_q^{\rm{R}}$, i.e.,
  \begin{align}
{\bf{G}} = \left[ {\begin{array}{*{20}{c}}
1&{{\bar T}}\\
0&1
\end{array}} \right] \otimes {{\bf{I}}_2},
{\bf{E}}_q^{\rm{R}} = \delta_q^{\rm R}\left[ {\begin{array}{*{20}{c}}
{\frac{1}{3}{\bar T}^3}&{\frac{1}{2}{\bar T}^2}\\
{\frac{1}{2}{\bar T}^2}&{\bar T}
\end{array}} \right] \otimes {{\bf{I}}_2},
  \end{align}
where $\bar T$ and $\delta_q^{\rm R}$ represent the time duration of each frame and the intensity of the Gaussian process noise, respectively.

Then, the delay and Doppler shift of the cascaded channel from AP $a'$ to AP $a$ via target $q$ in frame $n$ are  given by
\begin{align}
\tau_{aa'q}[n]
&=
\frac{1}{c_o}
\sum_{\bar a\in\{a,a'\}}
\sqrt{
\sum_{{\rm z}\in\{{\rm x},{\rm y}\}}
\left|
x_q^{{\rm R},{\rm z}}[n]
-
x_{\bar a}^{{\rm A},{\rm z}}
\right|^2
},
\\
f_{aa'q}[n]
&=
\sum_{\bar a\in\{a,a'\}}
\frac{
f_{\rm c}
\sum\limits_{{\rm z}\in\{{\rm x},{\rm y}\}}
v_q^{{\rm R},{\rm z}}[n]
\left(
x_q^{{\rm R},{\rm z}}[n]
-
x_{\bar a}^{{\rm A},{\rm z}}
\right)
}{
c_o
\sqrt{
\sum_{{\rm z}\in\{{\rm x},{\rm y}\}}
\left|
x_q^{{\rm R},{\rm z}}[n]
-
x_{\bar a}^{{\rm A},{\rm z}}
\right|^2
}
},
\end{align}
where $c_o$ is the speed of light and $f_{\rm c}$ is the carrier frequency.

\subsection{Elastic Network Topology (ENT)}
As illustrated in Fig. \ref{fig2Protocol}, an ENT-based service provisioning protocol is proposed, where each frame includes two phases:
\begin{itemize}
   \item  Phase-I: this phase operates exclusively within the intra-cell localized regime. All CCNs operate independently on their pre-assigned dedicated subbands and autonomously perform intra-cell local operations to classify the received service requests and partition resources between the local and federated regimes.
   \item Phase-II: this phase spans both the intra-cell localized regime and the inter-cell federated regime.
   Services and resources assigned to the local regime are managed locally by the DPU at the corresponding CCNs, while those assigned to the federated regime are grouped into $R$ CFNs and managed cooperatively by the CPU.
\end{itemize}
  
This two-phase dual-regime process adopts a partially cooperative service provisioning mechanism, which balances the performance gains of federated coordination against the reduced signaling overhead enabled by localized operation.\footnote{In this framework, the Phase-I decisions directly determine the feasible resource pool for Phase-II provisioning, and thus the Phase-II performance depends on the quality of the Phase-I decisions. Rather than guaranteeing instantaneous optimality for each phase at every frame, the proposed framework aims to achieve a practical balance between long-term utility and signaling overhead through sequential distributed decision-making.}

\begin{remark}
Throughout this paper, superscripts ``$\rm L$'' and ``$\rm F$''   denote variables associated with the localized regime ($\rm L$-regime) and federated regime ($\rm F$-regime), respectively.
\end{remark}

\subsection{Procedures in Phase-I}\label{sec2subC}
During this phase, the system operates exclusively within the L-regime.
Specifically, each CCN operates independently to perform the following procedures: local CSI estimation, local CSI exchange, local optimization, and result reporting.
Without loss of generality,  CCN $m$  is assumed to be operated on subband $m$. The detailed procedures are given~by:

\subsubsection{\textbf{Local CSI Estimation}} \label{sec2subC1}
In CCN $m$, all nodes ($k\in{\mathbb K}_m$ and $q\in{\mathbb Q}_m$) send service requests to the host AP of CCN~$m$.
Then, the users in $ {\mathbb K}_m$ transmit local pilot sequences in orthogonal time slots over all subcarriers on subband $m$  to APs in $  {\mathbb A}_m$.
Subsequently, each AP in $  {\mathbb A}_m$ estimates the corresponding CSI of these users on subband $m$.

\subsubsection{\textbf{Local CSI Exchange}} \label{sec2subC2}

All non-host APs of CCN $m$ forward their estimated CSI to the host AP of CCN $m$.

\subsubsection{\textbf{Local Optimization and Result Reporting}}\label{sec2subC3}

In  CCN $m$, the DPU executes the following three local procedures:

{\bf Service Classification}:
The requested services of CCN $m$ are dynamically classified into two categories: local services, provisioned in Phase-II within the L-regime using dedicated resources; and federated services, provisioned in Phase-II through inter-cell cooperation within the F-regime using integrated shared resources.
Specifically, the corresponding service index sets in CCN $m$ are given by:
\begin{subequations} \label{eqqos1}
\begin{align}
&\!\!\!\!\!{\rm{Local}}\;{\rm{service}}\;{\rm{set  }} : {\mathbb K}_m^{\rm{L}}[n] \cup {\mathbb Q}_m^{\rm{L}}[n],\label{eqqos1a}\\
&\!\!\!\!\!{\rm{Federated}}\;{\rm{service}}\;{\rm{set  }}: \hat {\mathbb K}_m^{\rm{F}}[n] \cup \hat {\mathbb Q}_m^{\rm{F}}[n],\label{eqqos1b}
\end{align}
\end{subequations}
where   ${\mathbb K}_m^{\mathrm{L}}[n]$ and ${\mathbb Q}_m^{\mathrm{L}}[n]$ denote the corresponding index sets of local communication and sensing services, respectively, while $\hat{\mathbb K}_m^{\mathrm{F}}[n]$ and $\hat{\mathbb Q}_m^{\mathrm{F}}[n]$ denote the corresponding index sets of federated communication and sensing services. Note, these sets must satisfy the following constraint:
\begin{align}
\left\{ {\begin{array}{*{20}{l}}
{\mathbb Z}_m^{\rm{L}}\left[ n \right] \cup \hat {\mathbb Z}_m^{\mathop{\rm F}\nolimits} \left[ n \right] = {{\mathbb Z}_m},\\
{\mathbb Z}_m^{\rm{L}}\left[ n \right] \cap \hat {\mathbb Z}_m^{\mathop{\rm F}\nolimits} \left[ n \right] = \emptyset,
\end{array}} \right. \;{\rm{for}}\; {\mathbb Z} \in\left\{{\mathbb K}, {\mathbb Q}\right\}. \label{eqqos2}
\end{align}

{\bf Resource Partitions}: The radio resource pool that will be used for service provisioning in Phase II is optimized in this phase and partitioned into two segments: a dedicated resource pool for local service provisioning in the L-regime and a shared resource pool for federated service provisioning in the F-regime. In this paper, we focus on the frequency and power resources. Specifically, recalling that each subband includes $B$ subcarriers and denoting the maximum transmission power of each AP by $P_{\rm max}$, the amounts of dedicated and shared radio resources allocated to the L-/F-regime are defined as :
\begin{align}
\left\{ {B_m^{\rm{X}}[n],{{\left[ {P_a^{\rm{X}}\left[ n \right]} \right]}_{a \in {{\mathbb A}_m}}}} \right\}, \;{\rm{for}}\; { \rm X} \in\left\{{\rm  L}, {\rm  F}\right\},\label{eqqos21}
\end{align}
respectively, subject to the following resource constraints:
\begin{align}
B_m^{\rm{L}}[n] + B_m^{\rm{F}}[n] = B;P_a^{\rm{L}}[n] +P_a^{\rm{F}}[n] = {P_{{\rm{max}}}}, \label{eqqos3}
\end{align}
where $B_m^{\rm{X}}[n] $ and  $P_a^{\rm{X}}[n]$ denote the corresponding number of subcarriers and transmission power available in the $\rm X$-regime in Phase-II, respectively.

{\bf Local Optimization Result Reporting}: Given the dedicated frequency and power resources in \eqref{eqqos21}, the DPU in CCN~$m$ optimizes the transmit powers and beamforming vectors of the APs within CCN~$m$ to support local service provisioning for users in ${\mathbb K}_m^{\mathrm{L}}[n] \cup {\mathbb Q}_m^{\mathrm{L}}[n]$.
The optimized configuration is then forwarded to the non-host APs of CCN~$m$ via fronthaul links, thereby preparing ISAC signal generation for local service provisioning in Phase-II. Simultaneously, the DPU in CCN~$m$ reports the remaining federated service requests in $\hat{\mathbb K}_m^{\mathrm{F}}[n] \cup \hat{\mathbb Q}_m^{\mathrm{F}}[n]$, together with the estimated CSI on the $m$-th subband for users in $\hat{\mathbb K}_m^{\mathrm{F}}[n]$, to the CPU, thereby preparing for cooperative processing in Phase-II.

\subsection{Procedures in Phase-II}

In Phase-II, the system operates concurrently in both the L-regime and the F-regime for local and federated service provisioning, respectively. The main steps are given as follows.

\subsubsection{\bf{Local Service Provisioning in the L-Regime}}
For each  CCN $m$, the APs in $  {\mathbb A}_m$ operate on subband $m$  and utilize the dedicated bandwidth/power resources defined in  \eqref{eqqos21} to transmit the downlink ISAC signals, provisioning the local services indexed in ${\mathbb K}_m^{\rm{L}}[n] \cup {\mathbb Q}_m^{\rm{L}}[n]$ as defined in \eqref{eqqos1a}.

\subsubsection{\bf{Federated Service Provisioning in the F-Regime}} \label{sec2subD2}
The CPU orchestrates the distributed shared resources defined in \eqref{eqqos21} across multiple CCNs to efficiently deliver federated services through the following steps.

{\textbf{CCN Grouping}}:
    Based on the classified federated services in \eqref{eqqos1b} reported by the DPUs, the CPU groups all CCNs into $R$ virtual CFNs for optimization complexity and signaling overhead reduction. The CFN $r$ is represented by  the tuple $\left\{  {{\mathbb K}}_r^{\rm F}[n], {{\mathbb Q}}_r^{\rm F}[n],  {{\mathbb A}}_r^{\rm F}[n] \right\}$, where
\begin{subequations}
\begin{align}
{\mathbb K}_r^{\rm F}[n]
&= 
\bigcup\nolimits_{m \in {\mathbb S}_r^{\rm F}[n]}
\widehat{\mathbb K}_m^{\rm F}[n],\\
{\mathbb Q}_r^{\rm F}[n]
&= 
\bigcup\nolimits_{m \in {\mathbb S}_r^{\rm F}[n]}
\widehat{\mathbb Q}_m^{\rm F}[n],\\
{\mathbb A}_r^{\rm F}[n]
&= 
\bigcup\nolimits_{m \in {\mathbb S}_r^{\rm F}[n]}
{\mathbb A}_m.
\end{align}
\end{subequations} 
where ${\mathbb S}_r^{\rm F}[n]$ is the index set of CCNs, and equivalently subbands, grouped into CFN $r$, and we have the following constraint
\begin{align}\!
\left\{ {\begin{array}{*{20}{l}}\!\!
{ \cup _{r = 1}^R{\mathbb Z}_r^{\rm{F}}\left[ n \right] =  \cup _{m = 1}^M{\hat{\mathbb Z}_m^{\rm{F}}\left[ n \right]}}, \\
{{\mathbb Z}_r^{\rm{F}}\left[ n \right] \cap  { {\mathbb Z}_{r'}^{\rm{F}}\left[ n \right]} = \emptyset,\;{\rm{if}}\;r \ne r'},
\end{array}} \right.
\!{\rm{for}}\;{\mathbb Z} \in \left\{ {{\mathbb K},{\mathbb Q}} \right\}.\label{eqb12}
\end{align}
 
\textbf{Federated CSI Estimation}: In CFN~$r$, the CPU schedules the federated users to transmit pilot sequences in orthogonal time slots, enabling CSI estimation for the inter-cell users on the other shared subbands that belong to the common subband set of CFN~$r$.
This additional channel estimation is required because each CCN reports only its local CSI to the CPU and has no knowledge of the inter-cell channels on the remaining shared subbands.
 Moreover, this step is essential because the APs in ${\mathbb A}_r^{\rm{F}}\left[ n \right]$ collaboratively utilize the shared bandwidth/power resources provided by the CCNs in  ${\mathbb S}_r^{\rm F}[n]$, supporting federated service provisioning for the services indexed by ${\mathbb K}_r^{\rm{F}}\left[ n \right] $ and ${\mathbb Q}_r^{\rm{F}}\left[ n \right] $, which requires the CSI of all users on all subbands of ${\mathbb S}_r^{\rm F}[n]$.

{\textbf{Federated CSI Exchange and Reporting}}:
In CFN~$r$, the non-host APs forward the estimated CSI on the shared subbands to their corresponding host APs via fronthaul links. The host APs then aggregate the received CSI and relay it to the CPU through the backhaul links.

{\textbf{Federated  Optimization and Execution}}:
The CPU jointly optimizes the downlink beamforming and power allocation schemes for each CFN to serve its associated federated services. The optimized configurations are then broadcast to the APs within each CFN. Subsequently, all APs in CFN~$r$ collaboratively transmit downlink ISAC signals to support federated service provisioning.

 \begin{remark}
Throughout this paper, for notational simplicity, the frame index $n$ is included in the definition of a frame-dependent variable upon its first occurrence and omitted thereafter unless ambiguity arises.  
Moreover, to unify the notation, the index $d$ is used to denote the CCN index $m$ (with $1\le m\le M$) when ${\rm X}={\rm L}$ and the CFN index $r$ (with $1\le r\le R$) when ${\rm X}={\rm F}$. 
Accordingly, we define the following unified definitions for the CCN/CFN index, AP set, communication service set, sensing service set, and subband set associated with regime ${\rm X}$ in frame $n$:
\begin{align}
&\left(d, {\mathbb A}_d^{\rm X}, {\mathbb K}_d^{\rm X}, {\mathbb Q}_d^{\rm X}, {\mathbb S}_d^{\rm X}\right) \nonumber\\
\triangleq&
\begin{cases}
\left(m, {\mathbb A}_m, {\mathbb K}_m^{\rm L}[n], {\mathbb Q}_m^{\rm L}[n], \{m\}\right), 
& \text{if } {\rm X}={\rm L},\\
\left(r, {\mathbb A}_r^{\rm F}[n], {\mathbb K}_r^{\rm F}[n], {\mathbb Q}_r^{\rm F}[n], {\mathbb S}_r^{\rm F}[n]\right), 
& \text{if } {\rm X}={\rm F}.
\end{cases}
\label{eq13}
\end{align} 
\end{remark}

\subsection{ISAC Signal Model}

This part introduces the signal model in Phase-II. Specifically, the downlink ISAC signals transmitted by AP $a$ over the $(m,b,l)$-th subband–subcarrier–time resource bin (RB) in frame $n$ under the $\mathrm{X}$-regime ($m\in{\mathbb S}_d^{\rm X}$, $\mathrm{X} \in \{\mathrm{L}, \mathrm{F}\}$) are
\begin{align}
{\bf{s}}_{a,mbl}^{\rm{X}}[ n ] &= \sum\nolimits_{k \in {\mathbb K}_d^{\rm{X}} } {\sqrt {p_{akm}^{{\rm{X}},{\rm{C}}}} [n] {\bf{w}}_{akm}^{{\rm{X}},{\rm{C}}}[ n ]s_{k,mbl}^{{\rm{X}},{\rm{C}}}\left[ n \right]}  \nonumber \\
&+\sum\nolimits_{q \in {\mathbb Q}_d^{\rm{X}}} {\sqrt {p_{aqm}^{{\rm{X}},{\rm{R}}}}[n] {\bf{w}}_{aqm}^{{\rm{X}},{\rm{R}}}\left[ n \right]s_{aq,mbl}^{{\rm{X}},{\rm{R}}}[n]},
\label{eqISACsignal}
\end{align}
where ${p_{azm}^{\rm{X},\rm{Y}}}[n]$ and ${{\bf w}_{azm}^{\rm{X},\rm{Y}}}[n]$ denote the corresponding transmission power and unit-norm beamforming vector, respectively, associated with communication service $z \in \mathbb{K}_d^{\rm{X}}$ if ${\rm Y} = {\rm C}$, and with radar sensing service $z \in \mathbb{Q}_d^{\rm{X}}$ if ${\rm Y} = {\rm R}$.
Besides, $s_{k,mbl}^{\rm{X},\rm{C}}[n]$ and $s_{aq,mbl}^{\rm{X},\rm{R}}[n]$ represent the corresponding communication and sensing symbols, respectively, each with unit power. Moreover, based on the definitions in \eqref{eqqos21} and \eqref{eq13}, the power constraint for each service is
\begin{align}
p_{azm}^{{\rm X},{\rm Y}} \ge 0,
\quad
p_{azm}^{{\rm X},{\rm Y}} = 0,
\quad
\text{if } z\notin{\mathbb Z}_d^{{\rm X},{\rm Y}},
\label{eqqos4}
\end{align} 
where ${\mathbb Z}_d^{{\rm X},{\rm Y}}$ is defined as 
${\mathbb K}_d^{\rm X}$ for communication services when ${\rm Y}={\rm C}$, 
and as ${\mathbb Q}_d^{\rm X}$ for sensing services when ${\rm Y}={\rm R}$. Here, ${\bf 1}_{\{\cdot\}}$ denotes the indicator function, which equals one if the condition inside the braces is satisfied and zero otherwise.
 
Also, the total power constraint for each AP is given by 
\begin{align}
&\sum\nolimits_{m \in {\mathbb S}_d^{\rm X}} B_m^{\rm X}
\left(
\sum\nolimits_{k \in {\mathbb K}_d^{\rm X}} p_{akm}^{{\rm X},{\rm C}}
+
\sum\nolimits_{q \in {\mathbb Q}_d^{\rm X}} p_{aqm}^{{\rm X},{\rm R}}
\right)
\nonumber\\
&\qquad\qquad\qquad\qquad
\leq
{\bf 1}_{\{a \in {\mathbb A}_d^{\rm X}\}} P_a^{\rm X},
\;
\forall a, \forall d. \label{eqqos7}
\end{align}

\section{Performance Metric Derivations}\label{sec:Performance}

This section derives the communication and sensing performance under the X-regime and the corresponding overhead.

\subsection{ Communication Performance Derivation}
In this part, we derive the communication performance in the $\rm X$-regime within a unified mathematical expression.

As described in Fig. \ref{fig2Protocol}, Section~\ref{sec2subC1}, and Section~\ref{sec2subD2}, communication users transmit pilot sequences in orthogonal time slots to estimate the CSI. Specifically, from \eqref{eqchannelN1} and given ${\bf{h}}_{akm}^{\rm{C}}[n-1]$, we know
\begin{align}
\!{\bf{h}}_{akm}^{\rm{C}}[n]\sim {\cal C}{\cal N}\left( {\rho _{ak}}{\bf{h}}_{akm}^{\rm{C}}[n - 1], \bar \lambda _{ak}^{\rm{C}}{\bf I}_{N_{\rm tx}} \right),
\end{align}
where $\bar \lambda _{ak}^{\rm{C}} = \left( {1 - \rho _{ak}^2} \right)\lambda _{ak}^{\rm{C}}$. 
Consequently, under the minimum mean square error (MMSE) estimation, the estimated channel of ${\bf{h}}_{akm}^{\rm{C}}[n]$ can be expressed as
\begin{align}
{\bf h}_{akm}^{\rm C}[n]
=
{\bf{\hat h}}_{akm}^{\rm C}[n]
+
{\bf e}_{akm}^{\rm C}[n],
\end{align}
where ${\bf{e}}_{akm}^{\rm{C}}[n]\sim {\cal C}{\cal N}\left( {\bf 0}_{N_{\rm tx}},\delta^{\rm C}_{akm}[n]{{\bf{I}}_{{N_{{\rm{tx}}}}}}\right)$ is the channel estimation error,  which is independent of the channel response. From \cite{ozdogan2018cell}, we know
 \begin{align}\label{eqa20}
\delta _{akm}^{\rm{C}}[n] = \bar \lambda _{ak}^{\rm{C}}\left[ {1 - \frac{{p_k^{{\rm{ce}}}D_{akm}^{{\rm{ce}}}[n]\bar \lambda _{ak}^{\rm{C}}}}{{p_k^{{\rm{ce}}}D_{akm}^{{\rm{ce}}}[n]\bar \lambda _{ak}^{\rm{C}} + \sigma _a^{\rm{A}}}}} \right],
\end{align}
where ${\sigma }_a^{\rm A}$ and  $p_k^{{\rm{ce}}}$   denote   the noise power at AP $a$ and the transmission power of uplink pilot symbol of user $k$, respectively.
Besides, $D_{akm}^{{\rm{ce}}}[n]$ denotes  the total number of pilot symbols for estimating channel $
 {\bf{\hat h}}_{akm}^{{\rm{C}}}[n]$, i.e.,
\begin{align}
D_{akm}^{\rm ce}[n]=
\begin{cases}
B\tilde D^{\rm ce},
& k\in{\mathbb K}_m,\\
B_m^{\rm F}\tilde D^{\rm ce},
& k\in{\mathbb K}_r^{\rm F}\setminus\widehat{\mathbb K}_m^{\rm F},\
m\in{\mathbb S}_r^{\rm F},
\end{cases}
\label{eq22}
\end{align}
where ${{{\tilde D}^{{\rm{ce}}}}}$  is the number of time slots of each user for channel estimation.
This is because, for  $ k \in {\mathbb K}_m$, $ {\bf{h}}_{akm}^{{\rm{C}}}[n]$  is estimated  over $B$ subcarriers on subband $m$ in the L-regime during Phase-I.  
In contrast, for 
$k\in{\mathbb K}_r^{\rm F}\setminus\widehat{\mathbb K}_m^{\rm F}$ and 
$m\in{\mathbb S}_r^{\rm F}$, 
${\bf h}_{akm}^{\rm C}$ is estimated over $B_m^{\rm F}$ subcarriers in the F-regime in Phase-II.

Upon transmitting the ISAC signals in \eqref{eqISACsignal} associated with the definition in \eqref{eq13}, the received  signal at user $k\in{\mathbb K}_d^{\rm X}$ in the $\rm X$-regime  over the ($m,b,l$)-th  subband-subcarrier-time RB is generally given by
 \begin{align}
&y_{k,mbl}^{{\rm{X}},{\rm{C}}}\left[ n \right] = \sum\nolimits_{a \in {{\mathbb A}_d^{\rm{X}}}} {{{\left( {{\bf{h}}_{akm}^{\rm{C}}} [n]\right)}^{\rm{H}}}} {\bf{s}}_{a,mbl}^{\rm{X}}[n] + u_{k,mbl}^{{\rm{X}},{\rm{C}}}\left[ n \right] \nonumber\\
&\!\!\!\!= \!  \sum\limits_{a \in {\mathbb A}_d^{\rm{X}}} {{{\left( {{\bf{e}}_{akm}^{\rm{C}}} \right)}^{\rm{H}}}}  {\bf{s}}_{a,mbl}^{\rm{X}} \!+\!\!\sum\limits_{a \in {\mathbb A}_d^{\rm{X}}} {{{\left( {{\bf{\hat h}}_{akm}^{\rm{C}}} \right)}^{\rm{H}}}} \!\!\!\sum\limits_{\bar k \in {\mathbb K}_d^{\rm{X}}} \!\!{\sqrt {p_{a\bar km}^{{\rm{X}},{\rm{C}}}} {\bf{w}}_{a\bar km}^{{\rm{X}},{\rm{C}}}s_{\bar k,mbl}^{{\rm{X}},{\rm{C}}}}  \nonumber\\
&+\!\!\sum\limits_{a \in {\mathbb A}_d^{\rm{X}}} {{{\left( {{\bf{\hat h}}_{akm}^{\rm{C}}} \right)}^{\rm{H}}}}\!\! \sum\limits_{q \in {\mathbb Q}_d^{\rm{X}}}\!\! {\sqrt {p_{aqm}^{{\rm{X}},{\rm{R}}}} {\bf{w}}_{aqm}^{{\rm{X}},{\rm{R}}}s_{aq,mbl}^{{\rm{X}},{\rm{R}}}} \! +\! u_{k,mbl}^{{\rm{X}},{\rm{C}}}[ n ],\label{eqq23}
   \end{align}
for $m\in{\mathbb S}_d^{\rm X}$, where  $u_{k,mbl}^{{\rm{X}},{\rm{C}}}\left[ n \right]  $ is the independent Gaussian noise with mean zero and covariance $\sigma_k^{\rm C}$.

Then, since the sensing signals are pre-determined sequences, the interference introduced by the second-to-last term on the right-hand side of \eqref{eqq23} can be effectively cancelled at the communication users through interference cancellation. As a result, the resulting SINR of user $k$ on subband $m$ in the $\rm X$-regime during frame $n$ is given by
\begin{align}
\Gamma _{km}^{{\rm{X}},{\rm{C}}}[n] &= \frac{{{{\left| {\sum\nolimits_{a \in {\mathbb A}_d^{\rm{X}}} {\sqrt{p_{akm}^{{\rm{X}},{\rm{C}}}}{{( {{\bf{\hat h}}_{akm}^{\rm{C}}} )}^{\rm{H}}}} {\bf{w}}_{akm}^{{\rm{X}},{\rm{C}}}} \right|}^2}}}{{{\sum\limits_{\bar k \in {\mathbb K}_d^{\rm{X}}\backslash \left\{ k \right\}}}{{\left| {\sum\limits_{a \in {\mathbb A}_d^{\rm{X}}} \!\!{\sqrt{p_{a\bar km}^{{\rm{X}},{\rm{C}}}}{{( {{\bf{\hat h}}_{akm}^{\rm{C}}} )}^{\rm{H}}}} \!\!{\bf{w}}_{a\bar km}^{{\rm{X}},{\rm{C}}}} \right|}^2} \!+ \!\sigma _{km}^{{\rm{X}},{\rm{C}}}[n]}}, \\
\sigma _{km}^{{\rm{X}},{\rm{C}}}[n]& = {\rm{E}}\left( {{{\left| {\sum\nolimits_{a \in {\mathbb A}_d^{\rm{X}}} {{{\left( {{\bf{e}}_{akm}^{\rm{C}}} \right)}^{\rm{H}}}} {\bf{s}}_{a,mbl}^{\rm{X}}} \right|}^2}} \right) + {\sigma _k^{\rm C}} \nonumber\\
& = \sum\limits_{a \in {\mathbb A}_d^{\rm{X}}} {\delta _{akm}^{\rm{C}}} \left(\! {\sum\limits_{\bar k \in {\mathbb K}_d^{\rm{X}}} {p_{a\bar km}^{{\rm{X}},{\rm{C}}}} \!  +\!  \sum\limits_{q \in {\mathbb Q}_d^{\rm{X}}} {p_{aqm}^{{\rm{X}},{\rm{R}}}} } \!\right)\! +\! \sigma_k^{\rm C},\label{eqb24}
\end{align}
where ${\delta _{akm}^{\rm{C}}}$ is defined in \eqref{eqa20}.

Therefore,  the effective transmission rate (nats/s/Hz) considering the signaling overhead of channel estimation for user $ k\in{\mathbb K}_d^{\rm X}$ in the X-regime during frame $n$  is
   \begin{align}
U_{n,k}^{{\rm X},{\rm C}}
=
\sum\nolimits_{m \in {\mathbb S}_d^{\rm X}}
\frac{B_m^{\rm X}L_m^{\rm X}}{MBL}
\ln\left(1+\Gamma_{km}^{{\rm X},{\rm C}}[n]\right).
\label{qosC}
\end{align}
where  $L$ denotes the total number of time slots in each frame, $B_m^{\rm{X}}$ is defined in \eqref{eqqos21}, and  $L_m^{\rm X}$  denotes the number of effective transmission time slots on subband $m$ in the X-regime, which is also defined as the remaining time slots after excluding the durations allocated for channel estimation, information/signaling exchange, and other related processes. The detailed calculation is provided in the following remark.
\begin{remark}
 As illustrated in Fig. \ref{fig2Protocol}, the dominant time consumption in the ENT arises from two parts: channel estimation and ISAC signal transmission on both the L-regime and F-regime. For simplicity, the time consumption of other procedures is neglected. Moreover, with the number of pilots defined in \eqref{eq22}, we have:
\begin{itemize}
      \item  For user $k\in{\mathbb K}_m^{\rm L}$ served  in the L-regime, the channel estimation consumes $K{{{\tilde D}^{{\rm{ce}}}}}$ time slots in Phase-I, leaving $L_m^{\rm L}=(L-K{{{\tilde D}^{{\rm{ce}}}}})$ time slots for ISAC signal transmission during Phase-II.
     \item For user $k\in{\mathbb K}_r^{\rm F}$ served  in the F-regime, CFN $r$
 incurs additional channel estimation overhead beyond the  $K{{{\tilde D}^{{\rm{ce}}}}}$  time slots consumed in Phase-I. Since each CFN must wait for the completion of channel estimation for all users across the shared subbands, the overhead is dominated by the subband with the largest number of unestimated users requiring CSI estimation, denoted by
$\tilde D^{\rm ce}\max_{j\in{\mathbb S}_r^{\rm F}}
\left|{\mathbb K}_r^{\rm F}\setminus\widehat{\mathbb K}_j^{\rm F}\right|$.
Therefore, the corresponding effective transmission time slots for CFN $r$ on subband $m$ are
$
L_m^{\rm F}
=
L-\tilde D^{\rm ce}
\left(
\max_{j\in{\mathbb S}_r^{\rm F}}
\left|
{\mathbb K}_r^{\rm F}\setminus\widehat{\mathbb K}_j^{\rm F}
\right|
+K
\right).$
      \end{itemize}
\end{remark}

\subsection{Sensing Performance Derivation}

In this part, we derive the radar sensing performance in the $\rm X$-regime within a unified formulation.
To sense target position and velocity, the APs, within the same  CCN  $d$ in the $\rm L$-regime or CFN  $d$ in the $\rm F$-regime, cooperatively transmit the ISAC signals on the allocated subbands defined in \eqref{eqISACsignal}.
With orthogonal bandwidth utilization among different CCNs or CFNs, the APs in ${{\mathbb A}_d^{\rm{X}}}$ receive only the ISAC signals reflected by all targets on their allocated subbands in ${{\mathbb S}_d^{\rm{X}}}$.
Besides, we assume that the impact of the NLoS component is negligible due to significant attenuation in the round-trip sensing channel among distributed APs. Additionally, the reflection from communication users is ignored, as their radar cross section (RCS) is assumed to be negligible.
Furthermore, the targets are considered to be well-separated, allowing the interference from other targets in the sensing of target $q$ to be ignored.
Moreover, perfect self-interference cancellation (SIC) is assumed for the full-duplex sensing operation at each AP.\footnote{
Note that NLoS components and residual SIC may increase the interference level, potentially degrading sensing accuracy and communication rate. However, such effects can be readily incorporated into the proposed framework by regarding it as additional interference terms.}

\begin{figure*}[ht]
\begin{subequations}\label{eqa35}
\begin{align}
{\left[ {{\bf{F}}_{aa'q}^{{\rm{X}},{\rm{P}}}} \right]_{1,1}} &= {\left( {\frac{{\partial {\tau _{aa'q}}}}{{\partial x_q^{{\rm{R}},{\rm{x}}}}}} \right)^2}\bar W_{aa'q}^{\tau \tau } + {\left( {\frac{{\partial {f_{aa'q}}}}{{\partial x_q^{{\rm{R}},{\rm{x}}}}}} \right)^2}\bar W_{aa'q}^{ff} - 2\frac{{\partial {f_{aa'q}}}}{{\partial x_q^{{\rm{R}},{\rm{x}}}}}\frac{{\partial {\tau _{aa'q}}}}{{\partial x_q^{{\rm{R}},{\rm{x}}}}}\bar W_{aa'q}^{\tau f},\\
{\left[ {{\bf{F}}_{aa'q}^{{\rm{X}},{\rm{P}}}} \right]_{1,2}} &= {\left[ {{\bf{F}}_{aa'q}^{{\rm{X}},{\rm{P}}}} \right]_{2,1}} = \frac{{\partial {\tau _{aa'q}}}}{{\partial x_q^{{\rm{R}},{\rm{x}}}}}\frac{{\partial {\tau _{aa'q}}}}{{\partial x_q^{{\rm{R}},{\rm{y}}}}}\bar W_{aa'q}^{\tau \tau } + \frac{{\partial {f_{aa'q}}}}{{\partial x_q^{{\rm{R}},{\rm{x}}}}}\frac{{\partial {f_{aa'q}}}}{{\partial x_q^{{\rm{R}},{\rm{y}}}}}\bar W_{aa'q}^{ff} - \left( {\frac{{\partial {f_{aa'q}}}}{{\partial x_q^{{\rm{R}},{\rm{x}}}}}\frac{{\partial {\tau _{aa'q}}}}{{\partial x_q^{{\rm{R}},{\rm{y}}}}} + \frac{{\partial {f_{aa'q}}}}{{\partial x_q^{{\rm{R}},{\rm{y}}}}}\frac{{\partial {\tau _{aa'q}}}}{{\partial x_q^{{\rm{R}},{\rm x}}}}} \right)\bar W_{aa'q}^{\tau f},\\
{\left[ {{\bf{F}}_{aa'q}^{{\rm{X}},{\rm{P}}}} \right]_{2,2}}& = {\left( {\frac{{\partial {\tau _{aa'q}}}}{{\partial x_q^{{\rm{R}},{\rm{y}}}}}} \right)^2}\bar W_{aa'q}^{\tau \tau } + {\left( {\frac{{\partial {f_{aa'q}}}}{{\partial x_q^{{\rm{R}},{\rm{y}}}}}} \right)^2}\bar W_{aa'q}^{ff} - 2\frac{{\partial {f_{aa'q}}}}{{\partial x_q^{{\rm{R}},{\rm{y}}}}}\frac{{\partial {\tau _{aa'q}}}}{{\partial x_q^{{\rm{R}},{\rm{y}}}}}\bar W_{aa'q}^{\tau f},\\
{\bf{F}}_{aa'q}^{{\rm{X}},{\rm{PV}}}& = \left[ {\begin{array}{*{20}{c}}
{\frac{{\partial {f_{aa'q}}}}{{\partial v_q^{{\rm{R}},{\rm{x}}}}}\left( { - \bar W_{aa'q}^{\tau f}\frac{{\partial {\tau _{aa'q}}}}{{\partial x_q^{{\rm{R}},{\rm{x}}}}} + \bar W_{aa'q}^{ff}\frac{{\partial {f_{aa'q}}}}{{\partial x_q^{{\rm{R}},{\rm{x}}}}}} \right)}&{\frac{{\partial {f_{aa'q}}}}{{\partial v_q^{{\rm{R}},{\rm{y}}}}}\left( { - \bar W_{aa'q}^{\tau f}\frac{{\partial {\tau _{aa'q}}}}{{\partial x_q^{{\rm{R}},{\rm{x}}}}} + \bar W_{aa'q}^{ff}\frac{{\partial {f_{aa'q}}}}{{\partial x_q^{{\rm{R}},{\rm{x}}}}}} \right)}\\
{\frac{{\partial {f_{aa'q}}}}{{\partial v_q^{{\rm{R}},{\rm{x}}}}}\left( { - \bar W_{aa'q}^{\tau f}\frac{{\partial {\tau _{aa'q}}}}{{\partial x_q^{{\rm{R}},{\rm{y}}}}} + \bar W_{aa'q}^{ff}\frac{{\partial {f_{aa'q}}}}{{\partial x_q^{{\rm{R}},{\rm{y}}}}}} \right)}&{\frac{{\partial {f_{aa'q}}}}{{\partial v_q^{{\rm{R}},{\rm{y}}}}}\left( { - \bar W_{aa'q}^{\tau f}\frac{{\partial {\tau _{aa'q}}}}{{\partial x_q^{{\rm{R}},{\rm{y}}}}} + \bar W_{aa'q}^{ff}\frac{{\partial {f_{aa'q}}}}{{\partial x_q^{{\rm{R}},{\rm{y}}}}}} \right)}
\end{array}} \right].
\end{align}
\end{subequations}
\hrulefill \vspace{-0.3cm}
\end{figure*}
Then, when AP $a\in{\mathbb A}_d^{\rm X}$ uses the unit-norm receive 
beamforming vector ${\bf w}_{aq}^{\rm Rx}[n]$ for target $q$, i.e.,
$\|{\bf w}_{aq}^{\rm Rx}[n]\|^2=1$, the received sensing signal for target 
$q$ at AP $a$ in the $(m,b,l)$-th RB is given by
\begin{align}
&y_{aq,mbl}^{{\rm X},{\rm R}}[n]\!
=\!
u_{aq,mbl}^{{\rm X},{\rm R}}[n]
\!+\!\!
\sum_{a'\in{\mathbb A}_d^{\rm X}}
\!\!\left(e^{{\rm j}2\pi
\left(
lT\frac{f_{\rm c}^m}{f_{\rm c}}f_{aa'q}[n]
-
b\Delta_f\tau_{aa'q}[n]
\right)}\right.
\nonumber\\
&\left.
\underbrace{
\chi_{aa'q}^m[n]
\!\left({\bf w}_{aq}^{\rm Rx}[n]\right)^{\rm H}\!\!
{\bf v}\left(\theta_{aq}^{\rm R}[n]\right)
}_{\varpi_{aa'q}^m[n]}
\underbrace{
\!{\bf v}^{\rm H}\!\!\left(\theta_{a'q}^{\rm R}[n]\right)
{\bf s}_{a',mbl}^{\rm X}[n]
}_{\tilde{s}_{a'q}^{{\rm X},mbl}[n]}\!\!\right),
\label{eqsensing1}
\end{align}
for  $m\in{\mathbb S}_d^{\rm X}$, where $u_{aq,mbl}^{{\rm{X}},{\rm{R}}}[n]\sim{\cal CN}\left(0,\sigma_a^{\rm A}\right)$ denotes the equivalent received noise term at AP $a$ with power $\sigma_a^{\rm A}$;
$f_{\rm c}^m=f_{\rm c}+(m-1)B\Delta_f$ denotes the carrier frequency of the $m$-th subband;
$\theta_{aq}^{\rm R}
=
\arctan\!\left(
\frac{x_a^{{\rm A},{\rm y}}-x_q^{{\rm R},{\rm y}}}
{x_a^{{\rm A},{\rm x}}-x_q^{{\rm R},{\rm x}}}
\right)$  denotes the relative target angle;
$\chi _{aa'q}^m[ n]\sim {\cal CN}\left(0,{\lambda _{aa'q}^{\rm{R}}[ n ]}\right)$ represents the corresponding cascaded/round-trip channel fading coefficient,
and $\lambda _{aa'q}^{\rm{R}}[n] = \frac{G_{\rm R}N_{\rm tx}^2{c_o^2{\mkern 1mu} {\sigma _{{\rm{RCS}},q}}}}{{{{(4\pi )}^3}f_{\rm c}^2\prod\nolimits_{j \in \{ a,a'\} } {{{\left( {\sum\nolimits_{{\rm{z}} \in \{ {\rm{x}},{\rm{y}}\} } {|x_q^{{\rm{R}},{\rm{z}}} - x_j^{{\rm{A}},{\rm{z}}}{|^2}} } \right)}^2}} }}$.
\begin{theorem}\label{theorem1}
Based on \eqref{eqsensing1} and noting that the desired sensing parameters of target $q$  are represented by ${\bf x}_q^{\rm R}$  in \eqref{eq5}, the Fisher Information Matrix (FIM) regarding ${\bf x}_q^{\rm R}[n]$   is
\begin{align}
\!\!{\bf{F}}_q^{\rm{X}}[n] \!= \!\!\sum\limits_{a' \in {\mathbb A}_d^{\rm{X}}}{\sum\limits_{a \in {\mathbb A}_d^{\rm{X}}} \frac{1}{\sigma_{a}^{\rm A} }{\!\left[ {\begin{array}{*{5}{c}}
\!\!\!{{\bf{F}}_{aa'q}^{{\rm{X}},{\rm{P}}}}[n]&\!\!{{\bf{F}}_{aa'q}^{{\rm{X}},{\rm{PV}}}}[n]\\
\!\!\!{{{\left( {{\bf{F}}_{aa'q}^{{\rm{X}},{\rm{PV}}}} [n]\right)}^{\rm{T}}}}&\!\!{{\bf{F}}_{aa'q}^{{\rm{X}},{\rm{V}}}}[n]
\end{array}} \!\!\right]} }
,\label{eq17}
\end{align}
where ${\bf{F}}_{aa'q}^{{\rm{X}},{\rm{P}}}$ and ${\bf{F}}_{aa'q}^{{\rm{X}},{\rm{PV}}} $ are given in \eqref{eqa35}, and we have
\begin{align}
&{\bf{F}}_{aa'q}^{{\rm{X}},{\rm{V}}}  = \left[\!\! {\begin{array}{*{20}{c}}
\!\!{{{\left( {\frac{{\partial {f_{aa'q}}}}{{\partial v_q^{{\rm{R}},{\rm x}}}}} \right)}^2}\bar W_{aa'q}^{ff}}&\!\!\!{\frac{{\partial {f_{aa'q}}}}{{\partial v_q^{{\rm{R}},{\rm x}}}}\frac{{\partial {f_{aa'q}}}}{{\partial v_q^{{\rm{R}},{\rm y}}}}\bar W_{aa'q}^{ff}}\\
\!\!{\frac{{\partial {f_{aa'q}}}}{{\partial v_q^{{\rm{R}},{\rm x}}}}\frac{{\partial {f_{aa'q}}}}{{\partial v_q^{{\rm{R}},{\rm y}}}}\bar W_{aa'q}^{ff}}&\!\!{{{\left( {\frac{{\partial {f_{aa'q}}}}{{\partial v_q^{{\rm{R}},{\rm y}}}}} \right)}^2}\bar W_{aa'q}^{ff}}
\end{array}}\!\!\! \right],\\ 
&\bar W_{aa'q}^{\tau \tau } = \sum\limits_{m \in {\mathbb S}_d^{\rm{X}}} \!{\sum\limits_{b = 0}^{B_m^{\rm{X}} - 1} \!{\sum\limits_{l = 0}^{L_m^{\rm{X}} - 1} {8{{\left( {\pi b{\Delta _f}} \right)}^2}\lambda _{aa'q}^{\rm{R}}} } }\Gamma _{a'qm}^{{\rm{X}},{\rm{R}}}[n],  \label{eqWtt}\\
&\bar W_{aa'q}^{ff} = \!\sum\limits_{m \in {\mathbb S}_d^{\rm{X}}} \!{\sum\limits_{b = 0}^{B_m^{\rm{X}} - 1} \!\!{\sum\limits_{l = 0}^{L_m^{\rm{X}} - 1} \!\!{8{{\left( {\pi \frac{{f_{\rm{c}}^m}}{{{f_{\rm{c}}}}}lT} \right)}^2}\!\lambda _{aa'q}^{\rm{R}}} } }\Gamma _{a'qm}^{{\rm{X}},{\rm{R}}}[n],  \label{eqWff}\\
&\bar W_{aa'q}^{\tau f} = \sum\limits_{m \in {\mathbb S}_d^{\rm{X}}}\! \!{\sum\limits_{b = 0}^{B_m^{\rm{X}} - 1}\!\! {\sum\limits_{l = 0}^{L_m^{\rm{X}} - 1} \!{8{{ {\pi^2 blT{\Delta _f}} }}\frac{f_{\rm c}^m}{f_{\rm c}}\lambda _{aa'q}^{\rm{R}}} } }\Gamma _{a'qm}^{{\rm{X}},{\rm{R}}}[n], \label{eqWtf}\\
&\Gamma_{a'qm}^{{\rm X},{\rm R}}[n]
= 
\sum\nolimits_{k \in {\mathbb K}_d^{\rm X}}
p_{a'km}^{{\rm X},{\rm C}}
\left|
{\bf v}\!\left(\theta_{a'q}^{\rm R}\right)^{\mathsf H}
{\bf w}_{a'km}^{{\rm X},{\rm C}}
\right|^2
\nonumber\\
&\qquad\qquad+
\sum\nolimits_{q' \in {\mathbb Q}_d^{\rm X}}
p_{a'q'm}^{{\rm X},{\rm R}}
\left|
{\bf v}\!\left(\theta_{a'q}^{\rm R}\right)^{\mathsf H}
{\bf w}_{a'q'm}^{{\rm X},{\rm R}}
\right|^2 . 
\end{align}
\end{theorem} 
\begin{IEEEproof}
Please refer to Appendix~\ref{proofA0}.
\end{IEEEproof}

Combining \eqref{eq5}, we have the Bayesian estimation problem when the sensing service is provisioned in the $\rm X$-regime:
\begin{align}
\left\{ {\begin{array}{*{20}{l}}
{{\bf{x}}_q^{\rm{R}}\left[ n \right] = {\bf{Gx}}_q^{\rm{R}}\left[ {n - 1} \right] + {\bm{\varepsilon }}_q^{\rm{R}}\left[ n \right],}\\
{{\bf{\hat x}}_q^{\rm{X,R}}\left[ n \right] = {\bf{x}}_q^{\rm{R}}\left[ n \right] + {\bm{\epsilon }}_q^{\rm{X,R}}\left[ n \right],}
\end{array}} \right.
\end{align}
where ${\bm{\epsilon }}_q^{\rm{X,R}}\left[ n \right]$ is the measurement error of ${\bf{x}}_q^{\rm{R}}\left[ n \right] $ from the observed data in \eqref{eqsensing1}. Here, ${\bm{\epsilon }}_q^{\rm{X,R}}\left[ n \right]$ is approximated as Gaussian noise with mean zero and covariance ${\bm{\Sigma }}_q^{{\rm{X,R}}} =  {\rm diag}{{{( {{\bf{F}}_q^{{\rm{X}}}} )}^{ - 1}}} $.

Consequently, the Bayesian CRB provides a lower bound on the mean squared error (MSE) matrix in estimating ${\bf{x}}_q^{\rm{R}}\left[ n \right]$, which is  the inverse of the Bayesian FIM (BFIM) \cite{van2006bayesian}, i.e.,
\begin{align}
{\bf J}_q^{\rm X}[n]
=
\left(
{\bf E}_q^{\rm R}
+
{\bf G}{\bf J}_q^{-1}[n-1]{\bf G}^{\mathsf T}
\right)^{-1}
+
\left({\bf\Sigma}_q^{{\rm X,R}}[n]\right)^{-1}.
\end{align}
where
$\left({\bf E}_q^{\rm R}
+
{\bf G}{\bf J}_q^{-1}[n-1]{\bf G}^{\mathsf T}
\right)^{-1}$
and
$\left({\bf\Sigma}_q^{{\rm X,R}}[n]\right)^{-1}$
represent the information matrices derived from the prior prediction and the measurement of 
${\bf x}_q^{\rm R}[n]$, respectively.
Note that ${\bf{J}}_q[n-1] = {\bf{J}}_q^{\rm{L}}[n-1]$ if $q \in {\mathbb{Q}}_m^{\rm{L}}[n-1]$; otherwise, ${\bf{J}}_q[n-1] = {\bf{J}}_q^{\rm{F}}[n-1]$.
Therefore, the radar sensing errors regarding the position and velocity of target $q$ are given by
 \begin{align}\label{eqb37}
U_{n,q}^{{\rm X,RP}}
&=
\left[
\left({\bf J}_q^{\rm X}\right)^{-1}
\right]_{1,1}
+
\left[
\left({\bf J}_q^{\rm X}\right)^{-1}
\right]_{2,2},
\quad q\in{\mathbb Q}_d^{\rm X},\\
U_{n,q}^{{\rm X,RV}}
&=
\left[
\left({\bf J}_q^{\rm X}\right)^{-1}
\right]_{3,3}
+
\left[
\left({\bf J}_q^{\rm X}\right)^{-1}
\right]_{4,4},
\quad q\in{\mathbb Q}_d^{\rm X}. \label{eqb38}
\end{align}

\begin{remark}
It is worth noting that configuring the resource allocation to minimize sensing errors depends on the previously estimated information matrix ${\bf J}_q[n-1]$, the constant matrices ${\bf E}_q^{\rm R}$ and ${\bf G}$, the unknown target state ${\bf x}_q^{\rm R}[n]$, as well as other resource-related configuration variables.
To address the main challenge of the dependency on the unknown ${\bf{x}}_q^{\rm{R}}[n]$ during system resource configuration optimization, we approximate it using the predicted value based on the previously estimated state, i.e., $
{\tilde {\bf{x}}}_{q}^{\rm{R}}[n] = {\bf{G}} {\hat {\bf{x}}}_{q}^{{\rm R}}[n - 1]$
for the subsequent resource configuration, where   ${\hat {\bf{x}}}_{q}^{{\rm R}}[n - 1]= {\hat {\bf{x}}}_{q}^{{\rm L},{\rm R}}[n - 1]$ if $q \in {\mathbb{Q}}_m^{\rm{L}}[n-1]$; otherwise, ${\hat {\bf{x}}}_{q}^{{\rm R}}[n - 1]= {\hat {\bf{x}}}_{q}^{{\rm F},{\rm R}}[n - 1]$.

\end{remark}

\subsection{Signaling Overhead Analysis}
In this subsection, we analyze the overall signaling overhead in the proposed ENT architecture.
Note that the signaling overhead in this paper is dominated by the following two components:
1) CSI estimation overhead: refers to the overhead caused by transmitting pilot sequences between users and APs for CSI estimation.
2) Information exchange overhead: refers to the overhead caused by exchanging CSI among APs, DPUs, and the CPU for resource allocation and user service scheduling.
Other signaling overhead components are assumed to be negligible compared with these dominant terms and are therefore omitted for analytical tractability.\footnote{Here, the signaling overhead is modeled in a simplified manner to capture the relative coordination cost under different cooperation regimes. While more detailed modeling may change the exact overhead values, the proposed algorithmic procedure remains applicable.}

Moreover, the CSI estimation overhead has already been accounted for in \eqref{qosC} and \eqref{QoSS}, as it affects the effective ISAC signal transmission time.
Therefore, in the following analysis, we focus only on the CSI exchange overhead. In \cite{cho2012feedback}, the CSI exchange overhead can be quantified by the number of channel coefficients exchanged over the network. Based on this principle, we evaluate the CSI exchange overhead as follows.

\subsubsection{CSI Exchange  Overhead in Phase-I}

As shown in Fig. \ref{fig2Protocol} and Section \ref{sec2subC}, the CSI exchange overhead within CCN $m$ during Phase-I involves two procedures in the L-regime:

{\bf Local CSI Exchange}: In this step, the non-host APs report the estimated CSI of all communication users within CCN $m$  on subband $m$ (i.e., ${\bf{\hat h}}_{akm}^{{\rm{C}}}[n], \forall k \in {\mathbb K}_m$) to the corresponding host AP.
Consequently, given that CCN $m$ contains $A-1$ non-host APs, serves $K$ users, and each channel has a dimension of $N_{\rm tx}$, the corresponding CSI exchange overhead in this step of CCN $m$ over the fronthaul link during Phase-I is
\begin{align}O_{n,m}^{\rm{L},(1)} =   \left( {A - 1} \right)K{N_{{\rm{tx}}}}. \label{eqsignaloverhead1}
\end{align}

{\bf Local Optimization Result Reporting}:
In this step, the DPU at the host AP reports the CSI of federated users to the CPU. Specifically, this includes the CSI of federated communication users, i.e., ${\bf{\hat h}}_{akm}^{{\rm{C}}}[n]$, $\forall k \in {\hat{\mathbb K}}_m^{\rm F}[n]$, and the target information of sensing users, i.e.,
${\tilde {\bf{x}}}_{q}^{\rm R}[n]$ and ${\bf J}_q[n-1]$, 
$\forall q \in \widehat{\mathbb Q}_m^{\rm F}[n]$.
Therefore, the CSI exchange overhead in this step of CCN $m$ over the backhaul link can be expressed as
\begin{align}O_{n,m}^{\rm{L},(2)} =   {\left| {\widehat {\mathbb K}_m^{\rm{F}}} \right| } A{N_{{\rm{tx}}}}+20{\left| {\widehat {\mathbb Q}_m^{\rm{F}}} \right| },\label{eqsignaloverhead2}
\end{align}
where the factor 20 accounts for the reporting of 4 coefficients in ${\tilde {\bf x}}_{q}^{\rm R}[n]$ and 16 coefficients in ${\bf J}_q[n-1]$.

\subsubsection{CSI Exchange Overhead in Phase-II}

As shown in Fig. \ref{fig2Protocol} and Section \ref{sec2subC}, the CSI exchange within CFN $r$ in Phase-II occurs during the {\bf Federated CSI Exchange and Reporting} step. 
Specifically, in CFN $r$, each non-host AP transmits the estimated CSI of users over the shared subbands in the common subband set of CFN $r$ to the corresponding host AP, which then forwards the collected CSI to the CPU. 

Then, for CCN $m$ belonging to CFN $r$, i.e., $m\in{\mathbb S}_r^{\rm F}$, the number of users whose CSI needs to be estimated on subband $m$ is 
$\left|{\mathbb K}_r^{\rm F}\right|-\left|\widehat{\mathbb K}_m^{\rm F}\right|$, whereas the number of users whose CSI needs to be estimated on each subband $j\in{\mathbb S}_r^{\rm F}\backslash\{m\}$ is $\left|{\mathbb K}_r^{\rm F}\right|$. This is because the estimated CSI of the communication users in CCN $m$ on subband $m$, i.e., ${\bf{\hat h}}_{akm}^{\rm C}$ for all $k \in \widehat{\mathbb K}_m^{\rm F}$, 
has already been reported in Phase-I. Therefore, the CSI exchange overhead of CFN $r$ during Phase-II is 
\begin{align}
O_{n,r}^{\rm F} 
=&
\sum\nolimits_{m\in {\mathbb S}_r^{\rm F}}
\left(
\sum\nolimits_{j \in {\mathbb S}_r^{\rm F}}
\left|{\mathbb K}_r^{\rm F}\right|
-
\left|\widehat{\mathbb K}_m^{\rm F}\right|
\right)
\left(A-1\right)N_{\rm tx}
\nonumber\\
&+
\sum\nolimits_{m \in {\mathbb S}_r^{\rm F}}
\left(
\sum\nolimits_{j \in {\mathbb S}_r^{\rm F}}
\left|{\mathbb K}_r^{\rm F}\right|
-
\left|\widehat{\mathbb K}_m^{\rm F}\right|
\right)
A N_{\rm tx}
\nonumber\\
=&
\sum\nolimits_{m \in {\mathbb S}_r^{\rm F}}
\left(
\left|{\mathbb K}_r^{\rm F}\right|
\left|{\mathbb S}_r^{\rm F}\right|
-
\left|\widehat{\mathbb K}_m^{\rm F}\right|
\right)
(2A-1)N_{\rm tx},
\label{QoSS}
\end{align} 
where the first term in the first equality accounts for the reporting overhead incurred by the $(A-1)$ non-host APs when forwarding their estimated CSI to the host AP for all CCNs $m$ in CFN $r$, while the second term corresponds to the overhead incurred by the host AP when forwarding the collected CSI from all $A$ APs to the CPU for all CCNs $m$ in CFN $r$.

\section{Problem Formulation}\label{sec:Formulation}
In this paper, we design a utility-to-signaling ratio (USR) to quantify the tradeoff between sensing/communication utility and
signaling overhead, i.e.,
\begin{align}
{\rm USR}[n]\triangleq\frac{{\cal U}[n]}{{\cal S}[n]}, \label{equsr}
\end{align}
where ${\cal U}[n]$ denotes the total normalized sensing and communication utility in frame $n$, and  ${\cal S}[n]$ denotes the total normalized signaling overhead in frame $n$.

Specifically, the total normalized utility can be written as
\begin{align}
\!\! {\cal U}[n]\! =\! \sum\nolimits_{m = 1}^M {\left( {\sum\nolimits_{k \in {\mathbb K}_m^{\rm{L}}\left[ n \right]} {\bar U_{n,k}^{{\rm{L}},{\rm{C}}}}  + \sum\nolimits_{q \in {\mathbb Q}_m^{\rm{L}}\left[ n \right]} {\bar U_{n,q}^{{\rm{L}},{\rm{R}}}} } \right)} \nonumber \\
+\sum\nolimits_{r = 1}^R {\left( {\sum\nolimits_{k \in {\mathbb K}_r^{\rm{F}}\left[ n \right]} {\bar U_{n,k}^{{\rm{F}},{\rm{C}}}}  + \sum\nolimits_{q \in {\mathbb Q}_r^{\rm{F}}\left[ n \right]} {\bar U_{n,q}^{{\rm{F}},{\rm{R}}}} } \right)}, \label{eqb43}
\end{align} 
where the first and second terms on the right-hand side of \eqref{eqb43} represent the total normalized utility achieved in CCNs and CFNs, respectively.
Here, $\bar U_{n,k}^{{\rm{X}},{\rm{C}}}$ and $\bar U_{n,q}^{{\rm{X}},{\rm{R}}}$ denote the normalized communication and sensing utilities, respectively, which are obtained according to \eqref{qosC}, \eqref{eqb37}, and \eqref{eqb38}, i.e.,
\begin{align}
\bar U_{n,k}^{{\rm{X}},{\rm{C}}} &= {\left( {\min \left[ {\frac{{U_{n,k}^{{\rm{X}},{\rm{C}}} - U_k^{{\rm{C}},\min }}}{{U_k^{{\rm{C}},\max } - U_k^{{\rm{C}},\min }}},1} \right]} \right)^ + },\\
\bar U_{n,q}^{{\rm X},{\rm R}}
&=
\left(
\min\left\{
\bar U_{n,q}^{{\rm X},{\rm RP}},
\bar U_{n,q}^{{\rm X},{\rm RV}},
1
\right\}
\right)^+ ,
\end{align}
where ${\left( x \right)^ + } = \max \left( {0,x} \right)$, $\bar U_{n,q}^{{\rm{X}},{\rm{RP}}} = \frac{{{{\log }_{10}}U_{q,\max }^{{\rm{RP}}} - {{\log }_{10}}U_{n,q}^{{\rm{X}},{\rm{RP}}}}}{{{{\log }_{10}}U_{q,\max }^{{\rm{RP}}} - {{\log }_{10}}U_{q,\min }^{{\rm{RP}}}}}$ and $\bar U_{n,q}^{{\rm{X}},{\rm{RV}}} = \frac{{{{\log }_{10}}U_{q,\max }^{{\rm{RV}}} - {{\log }_{10}}U_{n,q}^{{\rm{X}},{\rm{RV}}}}}{{{{\log }_{10}}U_{q,\max }^{{\rm{RV}}} - {{\log }_{10}}U_{q,\min }^{{\rm{RV}}}}}$.
Besides,  ${U_k^{{\rm C},\max }}$ and ${U_k^{{\rm C},\min }}$ denote the maximum and minimum required rates for user $k$, respectively. Also,  ${{U}_{q,\max }^{{\rm{RP}}}}$ and ${{U}_{q,\max }^{{\rm{RV}}}}$ represent the maximum allowable errors for the position and velocity of target $q$, while ${{U}_{q,\min }^{{\rm{RP}}}}$ and ${{U}_{q,\min }^{{\rm{RV}}}}$ denote the corresponding minimum required errors, respectively.
Moreover, the base-10 logarithm is applied to compress the sensing error range and make the normalization more stable.

Next, with the signaling overhead defined in \eqref{eqsignaloverhead1}, \eqref{eqsignaloverhead2}, and \eqref{QoSS}, the total normalized signaling overhead is expressed as
\begin{align}
\!\!{\cal S}[n]\! =\!\! \frac{1}{o}\left( {\sum\limits_{m = 1}^M {\left( {{\bar O_m} \!+\! O_{n,m}^{{\rm{L,}}\left( 1 \right)} \!+\! O_{n,m}^{{\rm{L,}}\left( 2 \right)}} \right)} \! + \! \sum\limits_{r = 1}^R {O_{n,r}^{\rm{F}}} } \right),
\end{align}
where the normalization parameter $o$ controls the scaling of the total signaling overhead, and the baseline ${\bar O_m}$ captures the inherent signaling cost of basic CCN deployment.\footnote{Note that a larger normalization value $o$ reduces the relative impact of signaling overhead and leads to higher USR values, whereas a smaller $o$ increases the sensitivity of USR to signaling cost and results in a more conservative performance evaluation. Moreover, when ${\bar O_m}$ is larger, the additional signaling overhead incurred by joint CFN operation becomes relatively less significant, which encourages more cooperative resource allocation decisions.}

Next, to maximize the USR in \eqref{equsr}, we consider the joint design of network topology (including service classification and CCN grouping), power allocation, and beamforming: \begin{align}
{{\mathbb V}[n]} = \left\{ {\left\{ {{\mathbb V}_{n,m}^{{\rm{LCP}}},{\mathbb V}_{n,m}^{{\rm{LPB}}}} \right\}_{m = 1}^M,{\mathbb V}_n^{{\rm{FG}}},\left\{ {{\mathbb V}_{n,r}^{\rm{FPB}}} \right\}_{r = 1}^R} \right\},
\end{align}
where ${\mathbb V}_{n,m}^{{\rm{LCP}}}$ denotes the local optimization variable set for service classification and dedicated resource (power/bandwidth) partition (LCP) at CCN $m$;
  ${\mathbb V}_{n,m}^{{\rm{LPB}}}$ denotes the local optimization variable set for power and beamforming  allocation (LPB)  in CCN $m$;
  ${{\mathbb V}_n^{{\rm{FG}}}}$ denotes the federated optimization variable set for CCN grouping (FG) at CPU;
  ${\mathbb V}_{n,r}^{{\rm{FPB}}}$ denotes the federated optimization variable set for power and beamforming allocation (FPB) in  CFN $r$. Specifically, we have
\begin{align}
\!\! {\mathbb V}_{n,m}^{{\rm{LCP}}}\! = \!&\left\{ {\underbrace {{\mathbb K}_m^{\rm{L}}[n],{\mathbb Q}_m^{\rm{L}}[n]}_{{\rm{Classification}}},\underbrace {B_m^{\rm{L}}[n],{{\left\{ {P_a^{\rm{L}}[n]} \right\}}_{a \in {{\mathbb A}_m}}}}_{{\rm{Dedicated}}\;{\rm{resource}}\;{\rm{partition}}}} \right\}, \label{eqopt1}\\
\!\!\!\!{\mathbb V}_{n,m}^{{\rm{LPB}}} \! = \!& \left\{ {\underbrace {p_{akm}^{{\rm{L}},{\rm{C}}}[n],p_{aqm}^{{\rm{L}},{\rm{R}}}[n]}_{{\rm{Power}}\;{\rm{allocation}}},\underbrace {{\bf{w}}_{akm}^{{\rm{L}},{\rm{C}}}[n],{\bf{w}}_{aqm}^{{\rm{L}},{\rm{R}}}[n]}_{{\rm{Beamforming}}\;{\rm{vectors}}}} \right. \nonumber\\
\!\!\!\!&\qquad\qquad \quad\left. {\left| {\forall \left( {a,k,q} \right) \in \left( {{{\mathbb A}_m},{{\mathbb K}_m},{{\mathbb Q}_m}} \right)} \right.} \right\}, \label{eqopt2}\\
\!\!\!\!{{\mathbb V}_n^{{\rm{FG}}}} \! = \!& \underbrace {\left\{ {{\mathbb S}_r^{\rm{F}}[n]} \right\}_{r = 1}^R}_{{\rm{CCN}}\;{\rm{Grouping}}}, \label{eqopt3}\\
\!\!\!\!{\mathbb V}_{n,r}^{{\rm{FPB}}}\! = \! &\left\{ {\underbrace {p_{akm}^{{\rm{F}},{\rm{C}}}[n],p_{aqm}^{{\rm{F}},{\rm{R}}}[n]}_{{\rm{Power}}\;{\rm{allocation}}},\underbrace {{\bf{w}}_{akm}^{{\rm{F}},{\rm{C}}}[n],{\bf{w}}_{aqm}^{{\rm{F}},{\rm{R}}}[n]}_{{\rm{Beamforming}}\;{\rm{vectors}}}} \right. \nonumber\\
\!\!\!\!&\!\!\!\!\!\!\!\left. {\left| {\left( {a,k,q,m} \right) \in \left( {{\mathbb A}_r^{\rm{F}}[n],{\mathbb K}_r^{\rm{F}}[n],{\mathbb Q}_r^{\rm{F}}[n],{\mathbb S}_r^{\rm{F}}[n]} \right)} \right.} \right\}. \label{eqopt4}
\end{align}

Subsequently, we formulate the following  signaling-efficient optimization problem:
\begin{align}\label{eqP0}
\tag{\bf P0}\mathop {\max }\limits_{{\mathbb V}[n]} \;&
{\rm USR}[n]\triangleq\frac{{\cal U}[n]}{{\cal S}[n]},\\
{\rm s.t. }
&\left| {{\mathbb S}_r^{\rm{F}}[n]} \right| \le {M_{\max }}, \label{eqpoc1}\\
&\eqref{eqqos2} , \eqref{eqqos3} , \eqref{eqb12}, \eqref{eqqos4}, {\rm and}\; \eqref{eqqos7},
\end{align}
where constraint \eqref{eqpoc1} enforces that the maximum number of CCNs within each CFN must not exceed the predefined threshold ${M_{\max }}$.
In this problem, if all services are classified as federated and the number of CFNs is set to $R=1$ with ${M_{\max }}=M$, the optimization reduces to a fully centralized CFN architecture. Conversely, if all services are classified as local, the formulation corresponds to a traditional decentralized CCN.
By leveraging adaptive service classification and CFN grouping, the proposed framework enables a flexible balance between signaling overhead and system utility, thereby achieving enhanced system performance.

However, although a single centralized optimizer could jointly optimize the local and federated variable sets under ideal conditions, this approach is impractical in practice due to the causality constraints imposed by the two-phase transmission protocol and the inherently dual-regime operation, since the global CSI required for centralized optimization cannot be obtained before the optimization process is executed \cite{chen2024learning}. Furthermore, adopting a fully centralized approach would lead to substantially higher computational complexity.
This motivates the development of a sequential distributed optimization algorithm that relies on local observations while mitigating performance degradation caused by partial observability.

\section{Algorithm Development: MADRL} \label{sec:Solution}
In this section, we first reformulate problem \ref{eqP0} as a hierarchical MADRL task. We then develop a centralized training and decentralized execution (CTDE) paradigm to efficiently learn coordinated policies for the resulting partially observable sequential decision-making problem.

Specifically, to solve the sequential distributed optimization problem in \ref{eqP0}, we reformulate it as a multi-agent Markov decision process (MDP) \cite{song2025multi,mohajer2024dynamic} comprising $2M + R + 1$ intelligent agents. These agents are categorized into local and federated roles, each assigned specific coordination tasks across the two-phase transmission protocol illustrated in Fig.~\ref{fig2Protocol}. Specifically, the local agents include $M$ LCP agents responsible for service classification and local resource partitioning (including power and bandwidth allocation), as well as $M$ LPB agents for local power control and beamforming optimization. The federated agents consist of one FG agent tasked with dynamic grouping of CCNs into CFNs, along with $R$ FPB agents dedicated to federated power control and beamforming optimization across all cells in the same CFN. Mathematically, the MADRL can be modeled as the following tuple \cite{akbari2021age}:
\begin{align}
\left\langle {{\bm {\cal S}}, {\bm {\cal A}},   {\cal  R},{\bm {\cal P}}},\gamma \right\rangle,
\end{align}
where ${\bm {\cal S}}$ denotes the global state space encompassing channel conditions, user demands, and network topology; ${\bm {\cal A}}$ denotes the joint action space comprising discrete-continuous hybrid decisions across all agents; $ {\cal  R}$ represents the joint reward function balancing communication throughput, radar sensing accuracy, and energy efficiency; ${\bm{ \cal P}}$ denotes the state transition probability function conditioned on the current state and joint actions; and $\gamma \in [0,1)$ is the discount factor for future rewards.

\begin{figure*}
   \centering
   \includegraphics[width=0.75\textwidth]{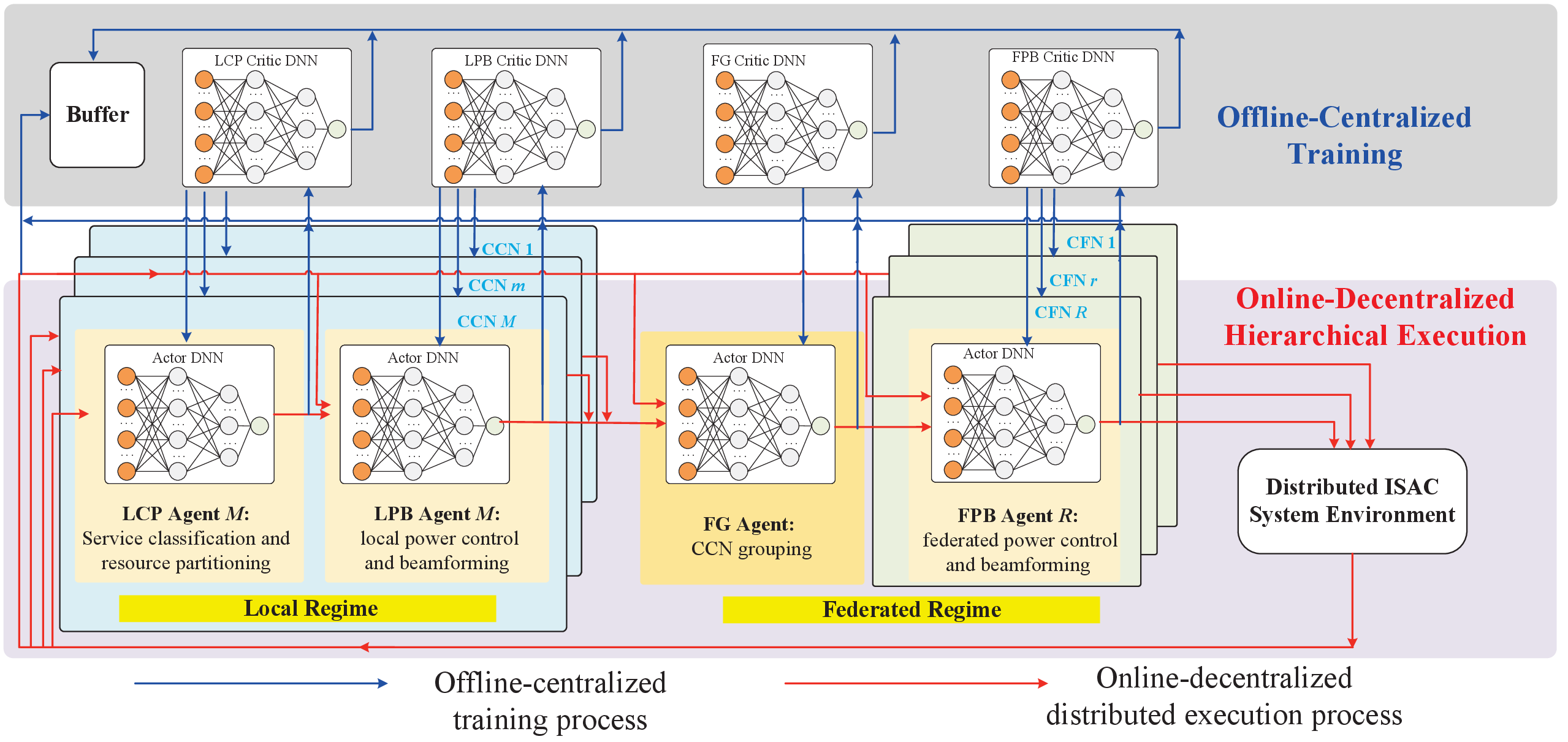}
   \caption{Illustration of the training and execution processes of the MAPPO algorithm for joint topology and resource optimization.}\label{fig3DRLAlgorithm}
 \end{figure*}

\subsection{Action}

In frame $n$, the joint action of all agents is denoted by $\boldsymbol{\alpha}_n \in  {\bm {\cal A}}$ and given by
\begin{align}
\boldsymbol{\alpha}_n
=
\left(
\{ \boldsymbol{\alpha}_n^{{\rm LCP}_m}, \boldsymbol{\alpha}_n^{{\rm LPB}_m} \}_{m=1}^{M},\;
\boldsymbol{\alpha}_n^{{\rm FG}},\;
\{ \boldsymbol{\alpha}_n^{{\rm FPB}_r} \}_{r=1}^{R}
\right),
\end{align}
where $\boldsymbol{\alpha}_n^{{\rm LCP}_m}$ and $\boldsymbol{\alpha}_n^{{\rm LPB}_m}$ denote the actions of the $m$-th LCP and LPB agents, respectively, while $\boldsymbol{\alpha}_n^{{\rm FG}}$ and $\boldsymbol{\alpha}_n^{{\rm FPB}_r}$, $1\le r\le R$, denote the actions of the FG agent and the $r$-th FPB agent, respectively. Note that the raw output of the deep neural network (DNN) deployed at each agent does not generally satisfy the constraints of problem ${\bf P0}$. Hence, to ensure that the joint action $\boldsymbol{\alpha}_n$ is feasible for ${\bf P0}$, the raw output is mapped to an executed action that satisfies the corresponding constraints.
Specifically, each DNN first produces a real-valued score vector, denoted by $\tilde{\boldsymbol{\alpha}}_n^{(\cdot)}$, and the corresponding feasible executed action $\boldsymbol{\alpha}_n^{(\cdot)}$ is obtained via the stage-dependent mapping
\begin{align}
\boldsymbol{\alpha}_n^{(\cdot)}
=
\mathcal{G}_{(\cdot)}^{\varsigma}
\!\left(
\tilde{\boldsymbol{\alpha}}_n^{(\cdot)};\tau_{(\cdot)}
\right), 
\end{align}
where $\varsigma={\rm tr}$ and $\varsigma={\rm de}$ denote the training and deterministic execution stages, respectively, and $\tau_{(\cdot)}$ denotes the corresponding temperature parameter.
In particular, the mapping $\mathcal{G}_{(\cdot)}^{\varsigma}(\cdot)$ is defined according to the type of decision variable. For a discrete decision induced by a score vector $\mathbf z\in\mathbb{R}^{L}$, we define
\begin{align}
\mathcal{G}_{{\rm cat}}^{\varsigma}(\mathbf z;\tau)
\triangleq
\begin{cases}
x,\; x \sim {\rm Cat}\!\left({\rm softmax}_{\tau}(\mathbf z)\right), & \varsigma={\rm tr},\\[1mm]
\displaystyle \arg\max_{\ell\in\{0,\ldots,L-1\}} z_{\ell+1}, & \varsigma={\rm de},
\end{cases}
\end{align}
which returns a label in $\{0,\ldots,L-1\}$. For a continuous share variable induced by $\mathbf z\in\mathbb{R}^{L}$, we define
\begin{align}
\mathcal{G}_{{\rm sha}}^{\varsigma}(\mathbf z;\tau)
\triangleq
\begin{cases}
{\rm Concrete}_{\tau}(\mathbf z), & \varsigma={\rm tr},\\
{\rm softmax}_{\tau}(\mathbf z), & \varsigma={\rm de}.
\end{cases}
\end{align}

Then, the details are provided as follows:
\subsubsection{LCP Agent Action}

The action of LCP$_m$ in frame $n$, i.e., $\boldsymbol{\alpha}_n^{{\rm LCP}_m}$, represents the executed values of the variables in $\mathbb V_{n,m}^{{\rm LCP}}$ defined in \eqref{eqopt1}. Let the raw DNN output of LCP$_m$ be
\begin{align}
\tilde{\boldsymbol{\alpha}}_n^{{\rm LCP}_m}
=
\left(
\{ \tilde{\boldsymbol{\pi}}_{n,k}^{{\rm C},m} \}_{k\in\mathbb K_m},
\{ \tilde{\boldsymbol{\pi}}_{n,q}^{{\rm R},m} \}_{q\in\mathbb Q_m},
\{ \tilde{\boldsymbol{\rho}}_{n,a}^{{\rm P},m} \}_{a\in\mathbb A_m},
\tilde{\boldsymbol{\rho}}_{n}^{{\rm B},m}
\right),
\end{align}
where $\tilde{\boldsymbol{\pi}}_{n,k}^{{\rm C},m}\in\mathbb{R}^{2}$ and $\tilde{\boldsymbol{\pi}}_{n,q}^{{\rm R},m}\in\mathbb{R}^{2}$ are the score vectors for service classification of communication user $k\in\mathbb K_m$ and radar target $q\in\mathbb Q_m$, respectively, $\tilde{\boldsymbol{\rho}}_{n,a}^{{\rm P},m}\in\mathbb{R}^{2}$ is the score vector for local/federated power partition of AP $a\in\mathbb A_m$, and $\tilde{\boldsymbol{\rho}}_{n}^{{\rm B},m}\in\mathbb{R}^{2}$ is the score vector for local/federated bandwidth partition of CCN $m$.

For $k\in\mathbb K_m$, $q\in\mathbb Q_m$, $a\in\mathbb A_m$, the corresponding local/federated service classification, power-partition share, and bandwidth-partition share are given by
\begin{align}
\pi_{n,k}^{{\rm C},m}
&=
\mathcal{G}_{{\rm cat}}^{\varsigma}
\!\left(
\tilde{\boldsymbol{\pi}}_{n,k}^{{\rm C},m};\tau_{\rm LCP}
\right),
\pi_{n,q}^{{\rm R},m}
=
\mathcal{G}_{{\rm cat}}^{\varsigma}
\!\left(
\tilde{\boldsymbol{\pi}}_{n,q}^{{\rm R},m};\tau_{\rm LCP}
\right), \nonumber\\
\boldsymbol{\rho}_{n,a}^{{\rm P},m}
&=
\mathcal{G}_{{\rm sha}}^{\varsigma}
\!\left(
\tilde{\boldsymbol{\rho}}_{n,a}^{{\rm P},m};\tau_{\rm LCP}
\right),
\boldsymbol{\rho}_{n}^{{\rm B},m}
=
\mathcal{G}_{{\rm sha}}^{\varsigma}
\!\left(
\tilde{\boldsymbol{\rho}}_{n}^{{\rm B},m};\tau_{\rm LCP}
\right).
\end{align}

Accordingly, the executed action $\boldsymbol{\alpha}_n^{{\rm LCP}_m}$ is given by
\begin{align}
{\mathbb K}_m^{\rm L}[n]
&=
\left\{
k \,\middle|\, \pi_{n,k}^{{\rm C},m}=0,\; k\in\mathbb K_m
\right\},\nonumber\\
{\mathbb Q}_m^{\rm L}[n]
&=
\left\{
q \,\middle|\, \pi_{n,q}^{{\rm R},m}=0,\; q\in\mathbb Q_m
\right\},\nonumber\\
P_a^{\rm L}[n]
&=
P_{\rm max}\,\boldsymbol{\rho}_{n,a}^{{\rm P},m}(1),
B_m^{\rm L}[n]
=
\left\lceil
B\,\boldsymbol{\rho}_{n}^{{\rm B},m}(1)
\right\rceil.
\end{align}
Note that $P_a^{\rm F}[n]$  and $ B_m^{\rm F}[n]$ can be determined once $P_a^{\rm L}[n]$  and $ B_m^{\rm L}[n]$ are known, according to \eqref{eqqos3}.

\subsubsection{LPB Agent Action}

The action of LPB$_m$ in frame $n$, i.e., $\boldsymbol{\alpha}_n^{{\rm LPB}_m}$, corresponds to the executed values of the variables in $\mathbb V_{n,m}^{{\rm LPB}}$ defined in \eqref{eqopt2}. Let the raw output of LPB$_m$ be
\begin{align}
\tilde{\boldsymbol{\alpha}}_n^{{\rm LPB}_m}
=
\left(
\{ \tilde{\boldsymbol{\rho}}_{n,a}^{{\rm L},m} \}_{a\in\mathbb A_m},
\{ \tilde{\boldsymbol{\Theta}}_{n,a}^{{\rm L},m} \}_{a\in\mathbb A_m}
\right),
\end{align}
where $\tilde{\boldsymbol{\rho}}_{n,a}^{{\rm L},m}\in\mathbb{R}^{K+Q}$ is the score vector for local power allocation at AP $a\in\mathbb A_m$, and $\tilde{\boldsymbol{\Theta}}_{n,a}^{{\rm L},m}\in\mathbb{R}^{(K+Q)\times\Delta_\Theta}$ is the score matrix for beam selection from the codebook $\bm{\Theta}$. Here, $\bm{\Theta}=\{\theta_\ell=(-\frac{1}{2}+\frac{\ell}{\Delta_\Theta})\pi\}_{\ell=1}^{\Delta_\Theta}\subset[-\frac{\pi}{2},\frac{\pi}{2}]$ is the discrete beam-angle codebook, and $\Delta_\Theta$ is its cardinality.\footnote{
The performance loss due to angle discretization is closely related to the antenna array size \cite{ge2020deep}. With a larger number of antennas, the resulting beams become narrower, such that a sufficiently fine angular resolution can effectively approximate continuous beamforming designs with negligible performance degradation.}

Note that the local service sets $\mathbb K_m^{\rm L}[n]\cup\mathbb Q_m^{\rm L}[n]$ have already been determined by LCP$_m$, thus only the valid local pairs participate in local power allocation and beam selection. Specifically, for each AP $a\in\mathbb A_m$, we define the masked local power-allocation score vector $\bar{\boldsymbol{\rho}}_{n,a}^{{\rm L},m}\in\mathbb{R}^{K+Q}$ as
\begin{align}
\bar{\rho}_{n,a}^{{\rm L},m}(i)
=
\begin{cases}
\tilde{\rho}_{n,a}^{{\rm L},m}(i), 
& \text{if } \substack{1\le i\le K\\ (m-1)K+i \in \mathbb K_m^{\rm L}[n]},\\
\tilde{\rho}_{n,a}^{{\rm L},m}(i), 
& \text{if } \substack{K+1\le i\le K+Q\\ (m-1)Q+(i-K) \in \mathbb Q_m^{\rm L}[n]},\\
-\infty, & \text{otherwise}.
\end{cases}
\end{align}
Then, the executed local power-allocation share vector of AP $a$ is generated as
\begin{align}
\boldsymbol{\rho}_{n,a}^{{\rm L},m}
=
\mathcal{G}_{{\rm sha}}^{\varsigma}
\!\left(
\bar{\boldsymbol{\rho}}_{n,a}^{{\rm L},m};\tau_{\rm LPB}
\right),
\end{align}
which is normalized only over the valid local pairs, while the entries corresponding to invalid pairs are forced to zero.

Similarly, for beam selection, we define the masked beam-selection score matrix $\bar{\boldsymbol{\Theta}}_{n,a}^{{\rm L},m}\in\mathbb{R}^{(K+Q)\times \Delta_\Theta}$ such that only the rows corresponding to valid local pairs are retained. Then, the beam-selection index is given by
\begin{align}
\vartheta_{n,a,i}^{{\rm L},m}
=
\mathcal{G}_{{\rm cat}}^{\varsigma}
\!\left(
\bar{\boldsymbol{\Theta}}_{n,a}^{{\rm L},m}(i,:);\tau_{\rm LPB}
\right),
\end{align} 
for all $i$ satisfying $(m-1)K+i \in \mathbb K_m^{\rm L}[n]$ if $1\le i\le K$ and $(m-1)Q+(i-K) \in \mathbb Q_m^{\rm L}[n]$ if $K+1\le i\le K+Q$. The corresponding selected beam angle $\theta_{n,a,i}^{{\rm L},m}$ is the $(\vartheta_{n,a,i}^{{\rm L},m}+1)$-th entry of the discrete beam-angle codebook $\bm{\Theta}$.

Accordingly, the executed local power-allocation variables and beamforming vectors in $\boldsymbol{\alpha}_n^{{\rm LPB}_m}$ are given by
\begin{align}
p_{akm}^{{\rm L},{\rm C}}[n]
&=
\frac{P_a^{\rm L}[n]}{B_m^{\rm L}[n]}\,
\boldsymbol{\rho}_{n,a}^{{\rm L},m}(i), \nonumber\\
\mathbf{w}_{akm}^{{\rm L},{\rm C}}[n]
&=
e^{-{\rm j}\hat{\varphi}_{akm}}\,
\mathbf{v}\!\left(
\theta_{n,a,i}^{{\rm L},m}
\right).
\end{align}
where $k=(m-1)K+i \in \mathbb K_m^{\rm L}[n]$ for $1\le i\le K$, and $\hat{\varphi}_{akm}
=
\angle\!\left[
\left(\hat{\mathbf{h}}_{akm}^{\rm C}\right)^{\mathsf H}
\mathbf{v}\!\left(
\theta_{n,a,i}^{{\rm L},m}
\right)
\right]$
is the extracted phase used to perform coherent combining in \eqref{eqb24}. Besides, we have
\begin{align}
p_{aqm}^{{\rm L},{\rm R}}[n]
=
\frac{P_a^{\rm L}[n]}{B_m^{\rm L}[n]}\,
\boldsymbol{\rho}_{n,a}^{{\rm L},m}(i)
,\mathbf{w}_{aqm}^{{\rm L},{\rm R}}[n]
=
\mathbf{v}\!\left(
\theta_{n,a,i}^{{\rm L},m}
\right),
\end{align}
where  $
q=(m-1)Q+(i-K)\in \mathbb Q_m^{\rm L}[n]$ for  $1\le i-K\le Q$.

\subsubsection{FG Agent Action}

The action of the FG agent in frame $n$, i.e., $\boldsymbol{\alpha}_{n}^{{\rm FG}}$, represents the executed values of the variables in $\mathbb{V}_{n}^{{\rm FG}}$ defined in \eqref{eqopt3}. Let its raw DNN output  be
\begin{align}
\tilde{\boldsymbol{\alpha}}_{n}^{{\rm FG}}
=
\left\{
\tilde{\boldsymbol{\pi}}_{n,m}^{\rm F}
\right\}_{m=1}^{M},
\end{align}
where $\tilde{\boldsymbol{\pi}}_{n,m}^{\rm F}\in\mathbb{R}^{R}$ is the score vector used to determine the grouping decision of CCN $m$ over the $R$ candidate CFNs.

Accordingly, the cluster-selection index is given by
\begin{align}
\pi_{n,m}^{\rm F}
=
\mathcal{G}_{{\rm cat}}^{\varsigma}
\!\left(
\tilde{\boldsymbol{\pi}}_{n,m}^{\rm F};\tau_{\rm FG}
\right)+1,
1\le m\le M,
\end{align}
where $\pi_{n,m}^{\rm F}\in\{1,\ldots,R\}$ denotes the selected CFN index for CCN $m$.
Therefore, the executed action of $\boldsymbol{\alpha}_{n}^{{\rm FG}}$ is
\begin{align}
\mathbb{S}_r^{\rm F}[n]
=
\left\{
m \mid \pi_{n,m}^{\rm F}=r,\;1\le m\le M
\right\},
1\le r\le R.
\end{align}

Note that If the resulting grouping violates $|{\mathbb S}_r^{\rm F}[n]|\le M_{\max}$, 
a projection operator is applied to restore feasibility by moving the excessive CCNs from overloaded CFNs to underloaded feasible CFNs according to the  assignment scores.

\subsubsection{FPB Agent Action}  

The action of FPB$_r$ in frame $n$, i.e., $\boldsymbol{\alpha}_n^{{\rm FPB}_r}$, corresponds to the executed values of the variables in $\mathbb{V}_{n,r}^{{\rm FPB}}$ defined in \eqref{eqopt4}. Let the raw DNN output of FPB$_r$ be
\begin{align}
\tilde{\boldsymbol{\alpha}}_n^{{\rm FPB}_r}
=
\left(
\{ \tilde{\boldsymbol{\rho}}_{n,a}^{{\rm F},r} \}_{a\in\mathbb A},
\{ \tilde{\boldsymbol{\Theta}}_{n,a}^{{\rm F},r} \}_{a\in\mathbb A}
\right),
\end{align}
where $\tilde{\boldsymbol{\rho}}_{n,a}^{{\rm F},r} \in \mathbb{R}^{M^2K + M^2Q}$ denotes the score vector for federated power allocation at AP $a \in \mathbb{A}=\{1,\ldots,MA\}$, and $\tilde{\boldsymbol{\Theta}}_{n,a}^{{\rm F},r} \in \mathbb{R}^{(M^2K + M^2Q) \times \Delta_\Theta}$ denotes the corresponding score matrix for beam selection. The first $M^2K$ entries correspond to communication-related subband–service positions across $M$ subbands and $MK$ users, while the remaining $M^2Q$ entries correspond to sensing-related subband–service positions across $M$ subbands and $MQ$ sensing tasks.

Given $\mathbb S_r^{\rm F}[n]$,  $\mathbb K_r^{\rm F}[n]$, and $\mathbb Q_r^{\rm F}[n]$ determined by the preceding LCP and FG actions, AP $a$ only allocates power and selects beams over the valid subband--service entries. Specifically, the valid entry index set is defined as
\begin{align}
\mathbb J_r^{\rm F}[n]
&\triangleq
\left\{
(m-1)MK+k
\,\middle|\,
\substack{
1\le m\le M,\; m\in\mathbb S_r^{\rm F}[n],\\
k\in\mathbb K_r^{\rm F}[n]
}
\right\}\nonumber\\
&\cup
\left\{
M^2K+(m-1)MQ+q
\,\middle|\,
\substack{
1\le m\le M,\\ m\in\mathbb S_r^{\rm F}[n],\\
q\in\mathbb Q_r^{\rm F}[n]
}
\right\}.
\end{align}

Accordingly, for each AP $a\in\mathbb A$, the masked federated power allocation score vector $\bar{\boldsymbol{\rho}}_{n,a}^{{\rm F},r}\in\mathbb{R}^{M^2K +M^2Q}$ is 
\begin{align}
\bar{\rho}_{n,a}^{{\rm F},r}(i)
=
\begin{cases}
\tilde{\rho}_{n,a}^{{\rm F},r}(i), & i\in \mathbb J_r^{\rm F}[n],\\
-\infty, & \text{otherwise},
\end{cases}
\end{align}
for $i=1,\ldots,M^2(K+Q)$.
Then, the executed federated power-allocation share vector of AP $a$ is generated as
\begin{align}
\boldsymbol{\rho}_{n,a}^{{\rm F},r}
=
\mathcal{G}_{{\rm sha}}^{\varsigma}
\!\left(
\bar{\boldsymbol{\rho}}_{n,a}^{{\rm F},r};\tau_{\rm FPB}
\right),
\end{align}
which is normalized only over the valid entries, while the entries corresponding to invalid entries are forced to zero.

Similarly, the masked beam-selection score matrix is denoted by $\bar{\boldsymbol{\Theta}}_{n,a}^{{\rm F},r}\in\mathbb{R}^{(M^2K +M^2Q)\times\Delta_\Theta}$. The corresponding beam-selection index is given by
\begin{align}
\vartheta_{n,a,i}^{{\rm F},r}
=
\mathcal{G}_{{\rm cat}}^{\varsigma}
\!\left(
\bar{\boldsymbol{\Theta}}_{n,a}^{{\rm F},r}(i,:);\tau_{\rm FPB}
\right),
i\in\mathbb J_r^{\rm F}[n].
\end{align}
Consequently, the corresponding selected angle $\theta_{n,a,i}^{{\rm F},r}$ is the $(
\vartheta_{n,a,i}^{{\rm F},r}+1)$-th entry of $\bm{\Theta}$.

Accordingly, the executed federated power-allocation variables and beamforming vectors in $\boldsymbol{\alpha}_n^{{\rm FPB}_r}$ are given by
\begin{align}
p_{akm}^{{\rm F},{\rm C}}[n]
&=
\frac{P_a^{\rm F}[n]}{B_m^{\rm F}[n]}\,
\boldsymbol{\rho}_{n,a}^{{\rm F},r}(i), \nonumber\\
\mathbf{w}_{akm}^{{\rm F},{\rm C}}[n]
&=e^{-{\rm j}\tilde{\varphi}_{akm}}\,
\mathbf{v}\!\left(
\theta_{n,a,i}^{{\rm F},r}
\right),
\end{align}
where $m\in\mathbb S_r^{\rm F}[n]$, $k\in\mathbb K_r^{\rm F}[n]$, $1\le i=(m-1)MK+k\le M^2K$,  and
$
\tilde{\varphi}_{akm}
=
\angle\!\left[
\left(\hat{\mathbf h}_{akm}^{\rm C}\right)^{\mathsf H}
\mathbf v\!\left(
\theta_{n,a,i}^{{\rm F},r}
\right)
\right]
$. Besides, we have
\begin{align}
p_{aqm}^{{\rm F},{\rm R}}[n]
&=
\frac{P_a^{\rm F}[n]}{B_m^{\rm F}[n]}\,
\boldsymbol{\rho}_{n,a}^{{\rm F},r}(i),
\mathbf{w}_{aqm}^{{\rm F},{\rm R}}[n]
=
\mathbf{v}\!\left(
\theta_{n,a,i}^{{\rm F},r}
\right),
\end{align}
where $m\in\mathbb S_r^{\rm F}[n]$, $q\in\mathbb Q_r^{\rm F}[n]$, $M^2K+1\le i=M^2K+(m-1)MQ+q\le M^2(K+Q)$.

\subsection{State}
In frame $n$, the global state ${\bf s}_n$ is formed by aggregating the observations of all agents, i.e.,
\begin{align}
{{\bf{s}}_n} = \left( {\{ {\bf{s}}_n^{{\rm{LC}}{{\rm{P}}_m}},{\bf{s}}_n^{{\rm{LP}}{{\rm{B}}_m}}\} _{m = 1}^M,\;{\bf{s}}_n^{{\rm{FG}}},\{ {\bf{s}}_n^{{\rm{FP}}{{\rm{B}}_r}}\} _{r = 1}^R} \right),
\end{align}
where  ${\bf{s}}_n^{{\rm{LC}}{{\rm{P}}_m}}$ and ${\bf{s}}_n^{{\rm{LP}}{{\rm{B}}_m}}$ denote the local observed states of the $m$-th LCP and LPB agents, respectively. In addition, ${\bf{s}}_{n}^{\rm{FG}}$ and ${\bf{s}}_{n}^{{\rm{FPB}}_r}$ denote the observed states of the FG agent and the $r$-th FPB agent, respectively.
These states can be expressed as:
 \begin{align}
&{\bf s}_{n}^{{\rm LCP}_m} &=& \left\{ {{\bf{x}}_{k}^{\rm{C}}, {\bf{x}}_{a}^{\rm{A}},{\bf{\hat h}}_{akm}^{\rm{C}}[n],\delta _{akm}^{\rm{C}}[n],{\bf{\tilde x}}_q^{\rm{R}}[n],{{\bf{J}}_q}[n - 1]} \right.\nonumber\\
&&&\qquad\qquad\left. {\left| {a \in {{\mathbb A}_m},k \in {{\mathbb K}_m},q \in {{\mathbb Q}_m}} \right.} \right\}, \nonumber  \\
 &{\bf s}_{n}^{{\rm LPB}_m} &=&
\left[
{\bf s}_{n}^{{\rm LCP}_m},
{\boldsymbol{\alpha}}_{n}^{{\rm LCP}_m}
\right],\nonumber \\
&{\bf{s}}_{n}^{\rm{FG}}  &=& \left[ {{\bf{\hat s}}_{n,1}^{\rm{F}},\ldots,{\bf{\hat s}}_{n,M}^{\rm{F}}} \right], \nonumber \\
&{\bf{s}}_{n}^{{\rm FPB}_r} &=&\left[\left\{ {{\bf{\hat s}}_{{n,m}}^{\rm{F}}} \right\}_{m \in {\mathbb S}_r^{\rm{F}}[n]},{\boldsymbol{\alpha}}_{n}^{\rm{FG}}\right]. \nonumber 
\end{align}
Here,  ${\bf{\hat s}}_{{n,m}}^{\rm{F}}$ denotes the corresponding reported state by the DPU of CCN $m$ to the CPU in the F-regime for subsequent federated service provisioning, i.e.,
 \begin{align}
{\bf{\hat s}}_{{n,m}}^{\rm{F}} &= \left\{{\bf{x}}_{k}^{\rm{C}}, {\bf{x}}_{a}^{\rm{A}}, {{\bf{\hat h}}_{akm}^{\rm{C}}[n],\delta _{akm}^{\rm{C}}[n],{\bf{\tilde x}}_q^{\rm{R}}[n],{{\bf{J}}_q}[n - 1],} \right.   \nonumber \\
& P_a^{\rm{F}}[n],B_m^{\rm F}[n]\left. {\left| {a \in {{\mathbb A}_m},k \in \widehat {\mathbb K}_m^{\rm{F}}[n],q \in \widehat {\mathbb Q}_m^{\rm{F}}[n]} \right.} \right\}.
\end{align}

\subsection{Reward}
To enhance numerical stability and convert the fractional trade-off between utility and cost into an additive form, we reformulate the fractional objective in frame $n$ into an equivalent logarithmic ratio as follows:
\begin{align}
{ {\cal  R}_n} = \log {\cal U}\left[ n \right] - \log {\cal S}\left[ n \right].
 \end{align}

\subsection{Multi-Agent Proximal Policy Optimization (MAPPO)}

We develop a MADRL framework based on MAPPO under the centralized training and decentralized execution (CTDE) paradigm. 
MAPPO is particularly suitable for this setting, since the decision process involves multiple heterogeneous agents and a high-dimensional mixed discrete--continuous action space, which can easily destabilize conventional learning methods. 
By leveraging clipped surrogate objectives and entropy regularization, MAPPO improves training stability, while its factored multi-agent policy structure naturally supports the role-specific decisions in our architecture.

Moreover, under the CTDE paradigm, the role-specific critics are trained using enriched agent-dependent features, consisting of each agent’s local observed state, executed action, and the shared global reward signal ${\cal R}_n$, thereby providing more stable value estimation and improving inter-agent coordination.
In contrast, during decentralized execution, each actor makes decisions solely based on its own causal local observed state, e.g., ${\bf s}_n^{{\rm LCP}_m}$ and ${\bf s}_n^{{\rm LPB}_m}$ for the local agents, ${\bf s}_n^{\rm FG}$ for the FG agent, and ${\bf s}_n^{{\rm FPB}_r}$ for the $r$-th FPB agent. 
In this way, the proposed framework preserves distributed sequential decision-making while still benefiting from role-specific critic learning with a shared global reward.
In the following, we introduce the proposed MAPPO training procedure.
 
\subsubsection{Decentralized Execution Design}

For notational simplicity, we define the set of agent types and their corresponding index sets as
\begin{align}
&{\mathbb U} = \left\{ {\rm LCP, LPB, FG, FPB} \right\},\\
&{{\mathbb I}_{{\rm LCP}}} = {{\mathbb I}_{{\rm LPB}}} = \left\{ {1,\cdots,M} \right\},\\
&{{\mathbb I}_{{\rm FG}}} = \left\{0\right\}, \quad
{{\mathbb I}_{{\rm FPB}}} = \left\{ {1,\cdots,R} \right\}.
\end{align}

In decentralized execution, each agent uses its own DNN-based actor to parameterize a stochastic policy. The joint policy is factorized independently across all agents:
\begin{align}
{\pi _{\bm\phi}}\left( {{{\boldsymbol{\alpha}}_n}\left| {{{\bf{s}}_n}} \right.} \right) 
= \prod\nolimits_{u \in {\mathbb U}} 
\prod\nolimits_{i \in {{\mathbb I}_u}} 
{\pi _{{{\boldsymbol\phi}^{u_i}}}^{u_i}
\left( {{\boldsymbol{\alpha}}_n^{u_i}\left| {{\bf{s}}_n^{u_i}} \right.} \right)},
\end{align}
where $\boldsymbol{\phi}^{u_i}$ denotes the DNN parameters of agent $u_i$, and  
$\boldsymbol{\phi} = \left\{ \boldsymbol{\phi}^{u_i} \right\}_{u \in \mathbb{U}, i\in{{\mathbb I}_u}}$ 
collects the DNN parameters of all agents.

\subsubsection{Centralized Critic Design}

Centralized critics are used to estimate the expected future returns of agents, providing a global perspective to guide policy updates during training.

To improve credit assignment and learning efficiency while respecting the heterogeneous semantics of different agent roles, we adopt a role-specific centralized critic architecture. 
Specifically, each role $u\in\mathbb U$ is associated with a dedicated DNN-based critic, which is represented by the value function $V_{\boldsymbol\psi^u}^u(\cdot)$, where $\boldsymbol\psi^u$ denotes the corresponding DNN parameters. 
All agents of the same role share the critic parameters $\boldsymbol\psi^u$. 
Then, for an agent with role $u$ and index $i\in{\mathbb I}_u$, its value estimate is given by
\begin{align}
V_{\boldsymbol\psi^u}^u\left( {\bf z}_n^{u_i} \right),
\end{align}
where $
{\bf z}_n^{u_i} = \left[ {\bf{s}}_n^{u_i}, {\boldsymbol{\alpha}}_n^{u_i}, {\bf e}_i \right]$
is the input feature vector for the agent of type $u$ with local index $i$.
Here, ${\bf e}_i$ is a one-hot encoding of the local index $i$.

Next, during the training process, each critic network computes its own temporal-difference (TD) error and generalized advantage estimate (GAE) using the shared global reward signal ${\cal R}_n$. 
For a specific agent $u_i$ at time frame $n$ in  $N_{\rm T}$ frames, the TD error $\delta_n^{u_i}$ is calculated as
\begin{align}
\delta_n^{u_i} 
= { {\cal  R}_n} 
+ \gamma V_{\boldsymbol\psi_{\rm old}^{u}}^u\left( {{\bf{z}}_{n+1}^{u_i}} \right) 
- V_{\boldsymbol\psi_{\rm old}^{u}}^u\left( {{\bf{z}}_n^{u_i}} \right).
\end{align}
Here, $\gamma \in [0, 1)$ is the discount factor, which determines the present value of future rewards.
The TD error quantifies the difference between the TD target and the current value estimate, providing the essential signal for estimating the advantage function.

Subsequently, the GAE ${\hat A}^{u_i}_{n}$ is computed through a backward recursion along the trajectory, i.e.,
\begin{equation}
{\hat A}^{u_i}_{n} = \delta^{u_i}_{n} + \gamma {\dot \lambda } {\hat A}^{u_i}_{n+1}, 
\quad n = N_{\rm T}, \ldots, 1,
\end{equation}
where ${\hat A}^{u_i}_{N_{\rm T}+1} = 0$. 
Besides, ${\dot \lambda }  \in [0,1]$ is the GAE parameter, which controls the bias-variance trade-off in advantage estimation.
A higher ${\dot \lambda }$ increases the effective planning horizon, favoring long-term rewards at the cost of higher variance, whereas a lower ${\dot \lambda }$ shortens the horizon, favoring immediate rewards with reduced variance but increased bias.

Finally, to stabilize policy gradient updates, we standardize the computed advantages for each role $u$ as follows:
\begin{align}
\bar A_n^{u_i} &= \frac{\hat A_n^{u_i} - {\mu _u}}{{\sigma _u}},\\
{\mu _u} &= \frac{1}{{N_{\rm T}|{{\mathbb I}_u}|}}
\sum\nolimits_{i \in {{\mathbb I}_u}} 
\sum\nolimits_{n = 1}^{N_{\rm T}} {\hat A_n^{u_i}}, \\
{\sigma _u} &= 
\sqrt{
\frac{1}{{N_{\rm T}|{{\mathbb I}_u}|}}
\sum\nolimits_{i \in {{\mathbb I}_u}} 
\sum\nolimits_{n = 1}^{N_{\rm T}} 
{{\left( {\hat A_n^{u_i} - {\mu _u}} \right)}^2}
}.
\end{align}

\subsubsection{Actor Objective}

For each role $u\in\mathbb U$, the actor aims to maximize the expected return while encouraging exploration. Accordingly, the actor loss is defined as
\begin{equation}
{\cal L}_\pi ^u
= \frac{-1}{{N_{\rm T}|{{\mathbb I}_u}|}}
\sum\nolimits_{i \in {{\mathbb I}_u}} 
\sum\nolimits_{n = 1}^{N_{\rm T}} 
\left( {G_n^{u_i} + {\eta _H}H_n^{u_i}} \right),
\end{equation}
where $G_n^{u_i}$ is the PPO clipped surrogate objective, $H_n^{u_i}$ is the policy entropy promoting exploration, and $\eta_H \ge 0$ is the entropy coefficient weighting the exploration incentive. These terms are calculated as follows:
\begin{align}
G_n^{u_i} 
&= \min \left( 
\xi_n^{u_i}\bar A_n^{u_i},
{\rm clip}\left( \xi_n^{u_i}, 1 \pm \dot \varepsilon \right)
\bar A_n^{u_i}
\right),\\
\xi _n^{u_i}
&=  
\frac{
\pi _{{{\boldsymbol\phi}^{u_i}}}^{u_i}
({\boldsymbol{\alpha}}_n^{u_i}\mid {\bf{s}}_n^{u_i})
}{
\pi _{{{\boldsymbol\phi}_{\rm old}^{u_i}}}^{u_i}
({\boldsymbol{\alpha}}_n^{u_i}\mid {\bf{s}}_n^{u_i})
},\\
H_n^{u_i}
&= -\!\!\!\sum\limits_{\boldsymbol{\alpha}^{u_i}\in{\cal A}^{u_i}}\!\!\!
\pi _{{{\boldsymbol\phi}^{u_i}}}^{u_i}
\left( \boldsymbol{\alpha}^{u_i} \!\!\mid {\bf s}_n^{u_i} \right)
\log
\pi _{{{\boldsymbol\phi}^{u_i}}}^{u_i}
\left( \boldsymbol{\alpha}^{u_i} \!\mid {\bf s}_n^{u_i} \right).
\end{align}
Here, ${\cal A}^{u_i}$ denotes the action space of agent $u_i$. 
The operator ${\rm clip}\left( \xi_n^{u_i},1 \pm \dot \varepsilon \right)$ restricts the probability ratio within the interval $\left[ {1 -  \dot \varepsilon, 1 +  \dot \varepsilon} \right]$. 
Besides, $\xi_n^{u_i}$ is the probability ratio between the new and old policies, and the clipping parameter $\dot \varepsilon$ constrains the policy update to enhance training stability.

\begin{table*}[t] {\small
\centering
\caption{Main System and Training Parameters}
\label{tab:simparam}
\renewcommand{\arraystretch}{1.05}
\setlength{\tabcolsep}{6pt}
\begin{tabular}{l l  l l}
\hline
\textbf{Parameter} & \textbf{Value } &
\textbf{Parameter} & \textbf{Value } \\ \hline

Antenna elements per AP ($N_{\mathrm{tx}}$) & $4$  &
Carrier frequency ($f_{\mathrm{c}}$) & $5.89~\mathrm{GHz}$ \\
Communication / radar users per cell ($K,Q$) & $(4,4)$ &
Subcarrier spacing ($\Delta f$) & $156.25~\mathrm{kHz}$ \\
Subcarriers per subband ($B$) & $16$ &
Symbols per frame ($L$) & $100$ \\
Communication requirement ($U_k^{\min},U_k^{\max}$) & $(0.05,\,0.35)$~nats/Hz/s &
Noise power density ($N_0$) & $-174~\mathrm{dBm/Hz}$ \\
Position sensing requirement ($U_{q,\min}^{\mathrm{RP}},U_{q,\max}^{\mathrm{RP}}$) & $(5\times10^{-5},\,2)$ &
Transmit power per AP ($P_{\rm max}$) & $40~\mathrm{dBm}$\\
Velocity sensing requirement ($U_{q,\min}^{\mathrm{RV}},U_{q,\max}^{\mathrm{RV}}$) & $(5\times10^{-5},\,2)$ &
Rician $K$-factor ($\kappa$) & $4$ \\
Channel correlation coefficient ($\rho_{ak}$) & $0.98$ &
Radar velocity range ($v_{\min},v_{\max}$) & $[20,\,80]~\mathrm{m/s}$ \\
State-transition intensity ($\delta_q$) & $100$ &
Speed of light ($c_o$) & $3{\times}10^{8}~\mathrm{m/s}$ \\
Discount factor ($\gamma$) & $0.9$ &
GAE parameter ($\dot \lambda$) & $0.96$ \\
PPO clip ratio ($\dot \varepsilon$) & Linear anneal $[0.3,0.15]$ &
Value loss weight ($c_V$) & $0.5$ \\
Entropy coefficient ($\eta_H$) & Linear anneal $[10^{-3},10^{-4}]$ &
Actor / critic learning rates ($\alpha_{\rm Le}$) & $5{\times}10^{-5}$ \\
\hline
\end{tabular}
}
\end{table*}

\subsubsection{Critic Objective}

For each role $u\in\mathbb U$, the critic is trained to minimize the prediction error against a target return. 
For agent $u_i$, the target return is constructed as
\begin{align}
\hat  {\cal  R}_n^{u_i} 
= \hat A_n^{u_i} 
+ V_{\boldsymbol\psi_{\rm old}^{u}}^u({\bf{z}}_n^{u_i}),
\end{align}
where $V_{\boldsymbol\psi_{\rm old}^{u}}^u({\bf{z}}_n^{u_i})$ denotes the value estimate used during trajectory collection. 
This target combines the advantage estimate with the old value estimate, thereby providing an estimate of the cumulative return used to train the critic.
The critic is then updated by minimizing the smooth L1 loss against this target, which provides robustness against outliers in the value target estimates and promotes more stable training dynamics compared to the standard mean squared error (MSE) loss, i.e., 
\begin{equation}
{\cal L}_V^u 
= \frac{1}{{N_{\rm T}|{{\mathbb I}_u}|}}
\sum\nolimits_{i \in {{\mathbb I}_u}} 
\sum\nolimits_{n = 1}^{N_{\rm T}} 
{\rho _1}\left( 
V_{\boldsymbol\psi^u}^u\left( {{\bf{z}}_n^{u_i}} \right) 
- \hat  {\cal  R}_n^{u_i} 
\right),
\end{equation}
where $\rho_1(\cdot)$ denotes the smooth L1 loss function:
\begin{align}
{{\rho }_1}\left( x \right) = 
\left\{ 
\begin{array}{ll}
0.5{x^2}, & {|x| < 1,}\\
{|x| - 0.5,} & {{\rm otherwise}.}
\end{array} 
\right.
\end{align}

 \begin{algorithm}[t]{\small
\caption{MAPPO for ENT-based Distributed ISAC}
\label{alg:abstract}
\begin{algorithmic}[1]
\Require Discount factor $\gamma$, GAE parameter ${\dot \lambda}$, clipping parameter $\dot\varepsilon$, value weight $c_V$, entropy weight $\eta_H$, and learning rate $\alpha_{\rm Le}$
\State Initialize actor and critic parameters $\{{\boldsymbol\phi}^{u_i},{\boldsymbol\psi}^u\}_{u\in\mathbb U,\, i\in{\mathbb I}_u}$
\State Initialize environment $\mathcal{E}$ and  buffer $\mathcal{B}$

\For{each episode $e=1,\ldots,E_{\max}$}
    \State Set old parameters ${\boldsymbol\phi}_{\rm old}^{u_i}\leftarrow{\boldsymbol\phi}^{u_i}$ and 
    ${\boldsymbol\psi}_{\rm old}^{u}\leftarrow{\boldsymbol\psi}^{u}$
    \State Reset $\mathcal{E}$ and observe the initial states $\{\mathbf{s}_1^{u_i}\}_{u\in\mathbb U,\, i\in{\mathbb I}_u}$
    \For{each time step $n=1,\ldots,N_{\rm T}$}
        \State Each agent $u_i$ samples an action 
        $\boldsymbol{\alpha}_n^{u_i}\sim \pi_{{\boldsymbol\phi}^{u_i}}^{u_i}(\cdot|\mathbf{s}_n^{u_i})$
        \State Execute the joint action 
        $\boldsymbol{\alpha}_n=\{\boldsymbol{\alpha}_n^{u_i}\}_{u\in\mathbb U,\, i\in{\mathbb I}_u}$ in $\mathcal{E}$
        \State Observe next states $\{\mathbf{s}_{n+1}^{u_i}\}_{u\in\mathbb U,\, i\in{\mathbb I}_u}$ and reward ${\cal R}_n$
        \State Store 
        $\left(\{\mathbf{s}_n^{u_i}\},\{\boldsymbol{\alpha}_n^{u_i}\}, {\cal  R}_n,\{\mathbf{s}_{n+1}^{u_i}\}\right)$ 
        into $\mathcal{B}$
    \EndFor

    \State For $\forall u\in\mathbb U$ and  $\forall i\in{\mathbb I}_u$, compute $\delta_n^{u_i}$, $\hat A_n^{u_i}$, $\bar A_n^{u_i}$, and $\hat  {\cal  R}_n^{u_i}$

    \State For  $\forall u\in\mathbb U$, compute the actor loss $\mathcal L_{\pi}^u$ and critic loss $\mathcal L_{V}^u$

    \State Compute the total loss:
  $    {{\cal L}_{{\rm total}}} 
    = \sum\nolimits_{u \in {\mathbb U}} 
    \left( {{\cal L}_\pi ^u + {c_V}{\cal L}_V^u} \right)
   $
   
   \State Update all actor and critic parameters:
     \begin{align*}
    {\boldsymbol\phi}^{u_i} 
    &\leftarrow 
    {\boldsymbol\phi}^{u_i} 
    - \alpha_{\rm Le}
    \nabla_{{\boldsymbol\phi}^{u_i}}
    \mathcal L_{\rm total},\\
    {\boldsymbol\psi}^u 
    &\leftarrow 
    {\boldsymbol\psi}^u 
    - \alpha_{\rm Le}
    \nabla_{{\boldsymbol\psi}^u}
    \mathcal L_{\rm total}.
    \end{align*}
\EndFor
\end{algorithmic}}
\end{algorithm}

\begin{figure*}[t]
  \centering
    \begin{minipage}{.45\textwidth}
   \centering
   \includegraphics[width=\textwidth]{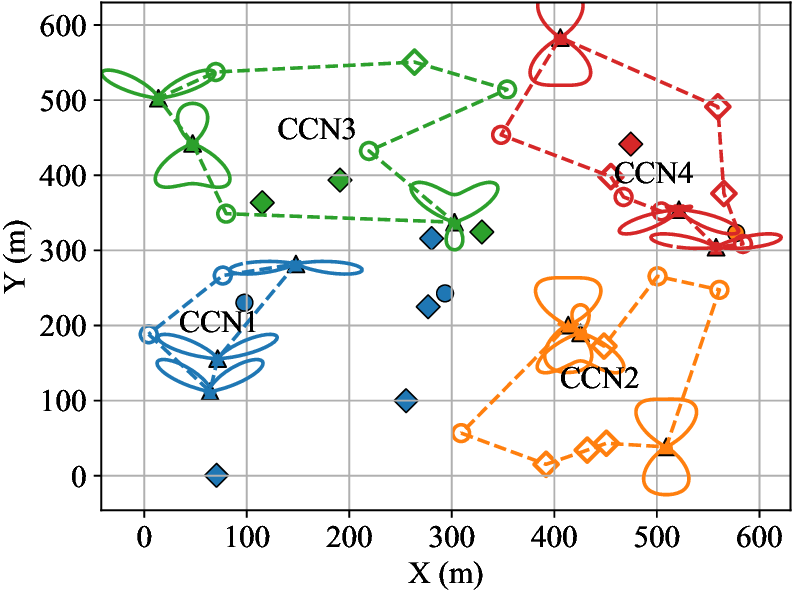}\vspace{-0.2cm}
   \caption*{  (a) Localized CCN Regime }\label{figsimu01}
  \end{minipage}
  \begin{minipage}{.45\textwidth}
   \centering
   \includegraphics[width=\textwidth]{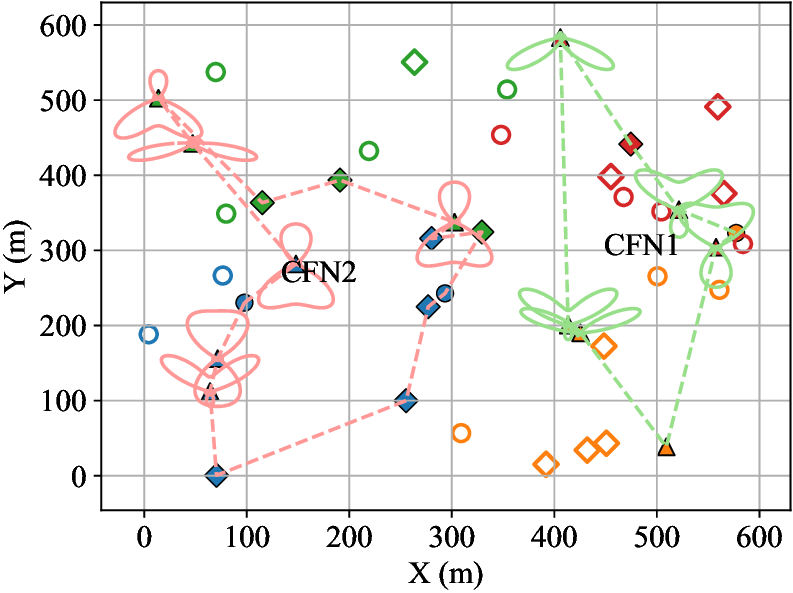}\vspace{-0.2cm}
    \caption*{  (b) Federated CFN Regime}\label{figsimu02}
  \end{minipage}
  \caption{Topology snapshots of the ENT-based distributed ISAC system in (a) the localized CCN regime and (b) the federated CFN regime.
Here, the triangle~($\blacktriangle$) denotes an AP, the hollow circle~($\circ$) denotes a local communication user, the hollow diamond~($\diamond$) denotes a local radar user, the solid circle~($\bullet$) denotes a federated communication user, and the solid diamond~($\blacklozenge$) denotes a federated radar user: $M=2{\times}2$, $R=2$, $M_{\rm max}=2$.}
\label{figsim0}
 \end{figure*}

\begin{figure*}[t]
  \centering
    \begin{minipage}{.3\textwidth}
   \centering
   \includegraphics[width=\textwidth]{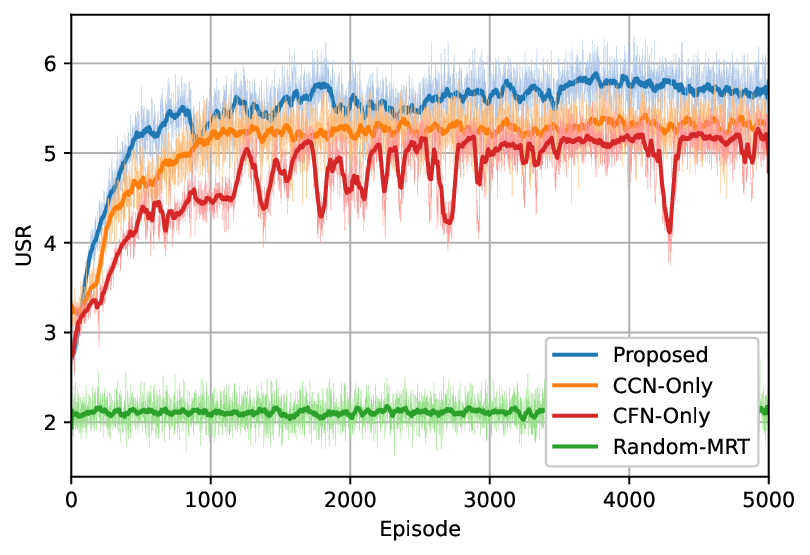}\vspace{-0.2cm}
    \caption*{(a) USR}\label{figsimu15}
  \end{minipage}
  \begin{minipage}{.3\textwidth}
   \centering
   \includegraphics[width=\textwidth]{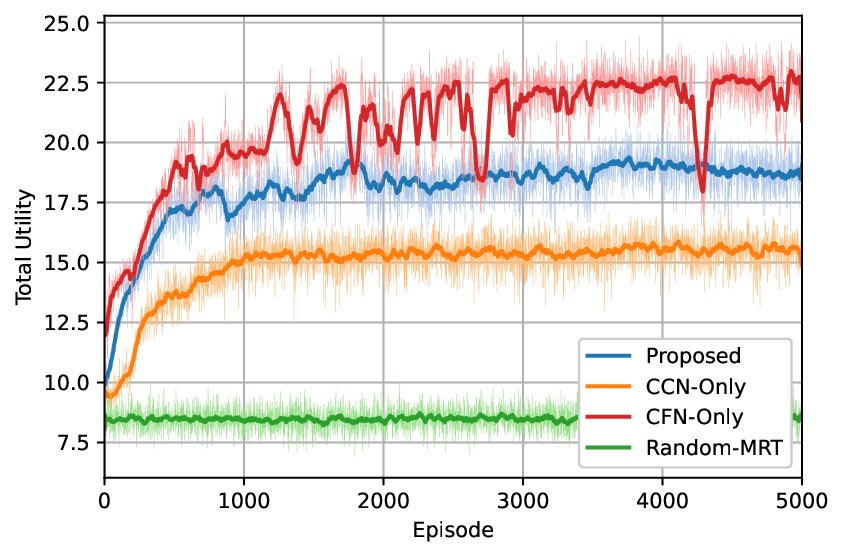}\vspace{-0.2cm}
   \caption*{(b) Total Utility }\label{figsimu11}
  \end{minipage}
\begin{minipage}{.3\textwidth}
   \centering
   \includegraphics[width=\textwidth]{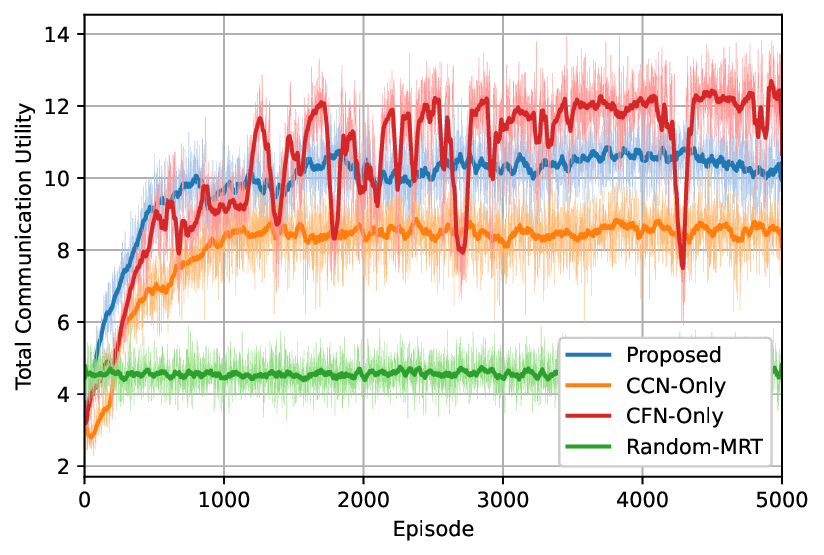}\vspace{-0.2cm}
    \caption*{(c)  Total Communication Utility }\label{figsimu13}
  \end{minipage}
 \begin{minipage}{.3\textwidth}
   \centering
   \includegraphics[width=\textwidth]{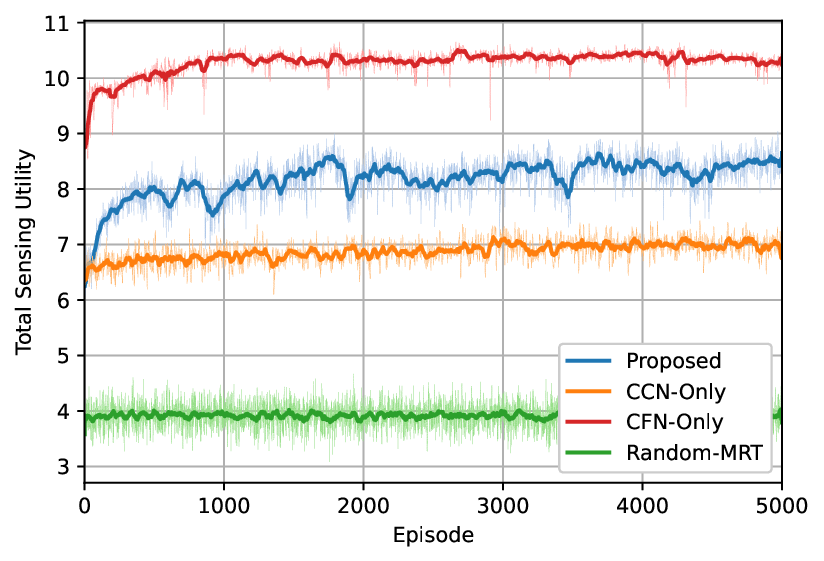}\vspace{-0.2cm}
    \caption*{(d) Total Sensing Utility}\label{figsimu14}
  \end{minipage}
  \begin{minipage}{.3\textwidth}
   \centering
   \includegraphics[width=\textwidth]{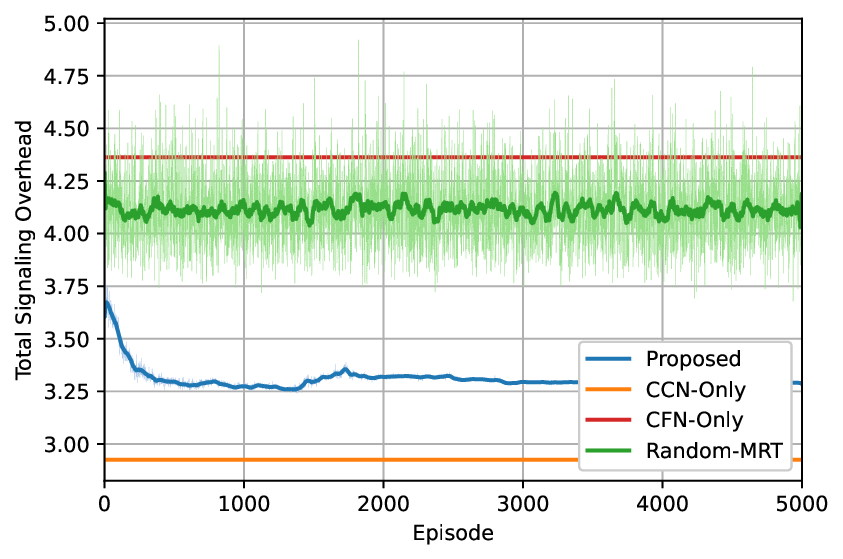}\vspace{-0.2cm}
    \caption*{(e) Total Signaling Overhead}\label{figsimu12}
  \end{minipage}
 \begin{minipage}{.3\textwidth}
   \centering
   \includegraphics[width=\textwidth]{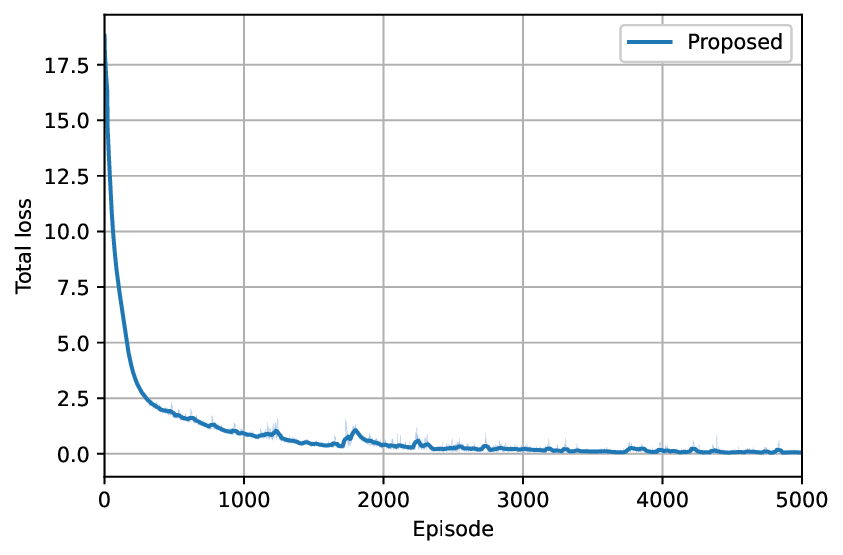}\vspace{-0.2cm}
    \caption*{(f) Loss}\label{figsimu16}
  \end{minipage}
  \caption{Performance comparisons among the proposed algorithm and other benchmarks with a moving average of 40 episodes: $M=2{\times}2$, $R=2$, $M_{\rm max}=2$. }\label{figsim1} \vspace{-0.3cm}
 \end{figure*}

\section{Simulation Results}\label{sec:Simulation}
The simulation results are presented to validate the effectiveness of the algorithm.
The network layout is illustrated in Fig.~\ref{figsim0}.
The simulation area covers $600{\times}600~\mathrm{m}^2$ and is divided into
$M$ CCNs, each of which contains $A=3$ APs.
To emulate spatial randomness and ensure topological diversity, all node positions are perturbed at the start of each episode by a random offset of up to $\pm1\%$ of the simulation area, while maintaining overall coverage stability.
The system operates on a 5.89 GHz carrier frequency from IEEE 802.11p \cite{nguyen2017delay}.
The path-loss function is defined by 
$\mathrm{PL}\!\left(\mathbf{x}_k^{\mathrm{C}},\mathbf{x}_a^{\mathrm{A}}\right)
= 32.4
+ 45\,\log_{10}\!\left(\frac{\|\mathbf{x}_k^{\mathrm{C}}-\mathbf{x}_a^{\mathrm{A}}\|_2}{1~\mathrm{m}}\right)
+ 20\,\log_{10}\!\left(\frac{f_{\mathrm{c}}}{1~\mathrm{GHz}}\right)\,[\mathrm{dB}]$.
The uplink transmit power and pilot length for channel estimation are set as   $p_k^{{\rm{ce}}}=25$ dBm and ${{\tilde D}^{{\rm{ce}}}}= 1$.
The signaling overhead parameters are set as ${o}=8M(A-1)N_{\rm tx}K$ and ${\bar O_m}=0.7o$.
Moreover, all actor and critic networks are implemented as multilayer perceptrons, where the critic adopts layer sizes of $[512,\,512,\,512]$ and the actor uses $[256,\,256,\,256,\,256]$. Both networks employ LeakyReLU activations with a negative slope of $0.01$ and apply layer normalization after each hidden layer. 
The temperature parameters are set to $\tau_{\rm LCP}=1.2$, $\tau_{\rm LPB}=0.3$, $\tau_{\rm FG}=1.0$, and $\tau_{\rm FPB}=0.8$, respectively. 
For comparison, the proposed framework is evaluated against the following baseline schemes:
\textbf{(i) CCN-Only:} The framework operates in each local CCN, where all users and resources are restricted to local allocation without any federated coordination. Here, only the LPB agents are trained.
\textbf{(ii) CFN-Only:} The framework operates in a fully federated regime, where all users and resources are centrally coordinated within the corresponding CFNs. Here, only the FG and FPB agents are trained.
\textbf{(iii) Random-MRT:} A benchmark in which each AP randomly classifies users into either local or federated service sets and performs maximum-ratio transmission (MRT) beamforming with equal power allocation.

From Fig.~\ref{figsim0}, we illustrate the topology snapshots of the ENT-based distributed ISAC system in both the L- and F-regimes.
It can be observed that radar users tend to prefer federated coordination since their interference footprint is smaller compared to communication users, allowing them to gain more performance benefits through joint beamforming.
In contrast, communication users already achieve satisfactory performance through local optimization within each CCN and therefore exhibit less incentive to participate in federated coordination. This adaptive clustering behavior highlights the algorithm's capability to balance sensing and communication objectives dynamically across the network. Moreover, the chosen beamforming vectors in both  L- and F-regimes are well aligned with the most effective users, demonstrating the effectiveness of the beamforming design.

\begin{figure}
   \centering
   \includegraphics[width=0.45\textwidth]{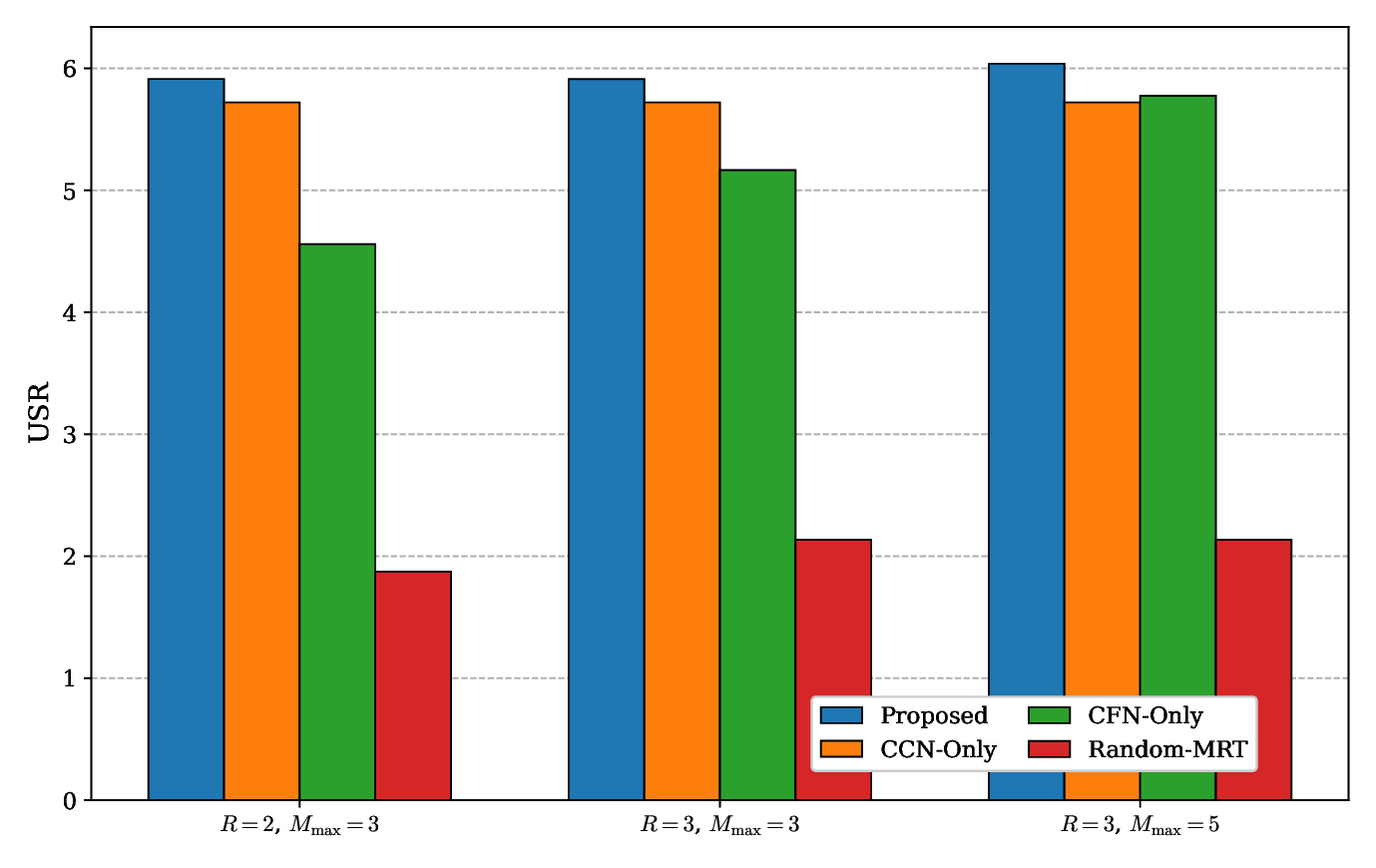}
   \caption{Performance comparisons among the proposed algorithm and other benchmarks with different grouping parameters: $M=3{\times}2$.}\label{fig4bar}
 \end{figure}

Fig.~\ref{figsim1} compares the proposed scheme with baseline methods across several performance metrics over training episodes, including:
(a) USR,
(b) total utility,
(c) total communication utility,
(d) total  sensing utility,
(e) total signaling overhead,
and (f)  training loss.
As shown in Figs.~\ref{figsim1}(a)–(d), the USR, total utility, total communication utility, and total sensing utility of all learning-based schemes increase steadily and eventually converge, confirming the convergence of the MAPPO framework. All three learning-based approaches significantly outperform the non-learning Random-MRT baseline in terms of USR, communication utility, and sensing utility, demonstrating the effectiveness of the proposed MAPPO-based design.
Among the learning-based schemes, the CFN-Only method achieves the highest communication and sensing utilities, as full federated cooperation provides the greatest spatial diversity gain.
However, Fig.~\ref{figsim1}(e) indicates that the CFN-Only method also incurs the highest signaling overhead due to the extensive CSI exchange required for federated cooperation. In contrast, the CCN-Only scheme has the lowest signaling overhead but also the lowest communication and sensing performance, as it avoids federated cooperation entirely.
By striking an effective balance between these trade-offs, the proposed scheme achieves the highest USR, as shown in Fig.~\ref{figsim1}(a). Specifically, it attains approximately a 10\% USR improvement compared with the CCN-Only method and about a 12\% improvement compared with the CFN-Only method after convergence. This result demonstrates the benefit of achieving a better compromise between service utility enhancement and signaling overhead, thereby outperforming both baselines.
Finally, Fig.~\ref{figsim1}(f) illustrates that the total training loss decreases monotonically and stabilizes quickly, further validating the convergence and numerical stability of the proposed framework.

Fig.~\ref{fig4bar} shows the performance comparison between the proposed algorithm and the benchmark schemes under different grouping parameters. It is observed that both the proposed algorithm and the CFN-Only scheme achieve better performance as the maximum number of clusters and the maximum group size increase, since the enlarged coordination space offers more degrees of freedom for cooperation. Moreover, the proposed algorithm consistently outperforms the baseline schemes, which demonstrates its effectiveness and robustness.

\section{ Conclusions}\label{sec:Conclusion}

In this paper, we propose an ENT-based network architecture designed to overcome the limitations of conventional CCN and CFN architectures in distributed ISAC networks.
By supporting autonomous CCN operations in the localized regime and dynamic CFN aggregation in the federated regime, this topology flexibly orchestrates the boundaries between localized CCNs and federated CFNs to balance signaling overhead and service performance, thereby enabling efficient and adaptive service provisioning.
To characterize this trade-off, we introduce the USR metric and formulate a USR-maximization problem that jointly optimizes service classification, network aggregation, and the allocation of power and spatial resources.
Considering the dual localized and federated operating regimes and the partially observable nature of distributed ISAC environments, we develop a MADRL framework based on MAPPO under the CTDE paradigm. Simulation results demonstrate that the proposed approach can effectively coordinate CCN and CFN operations and achieve efficient and high-quality ISAC service provisioning under dynamic network conditions.

\appendix

\subsection{Proof of Theorem \ref{theorem1}}\label{proofA0} 
We prove this theorem by first obtaining the expression of the FIM and then computing and simplifying its explicit form.

\subsubsection{FIM Expression}\label{proofA}
First, \eqref{eqsensing1} can be re-expressed as:
\begin{align}
 y_{aq,mbl}^{{\rm{X}},{\rm{R}}} =&
\sum\limits_{a' \in {{\mathbb A}_d^{\rm{X}}}} {\underbrace {\tilde s_{a'q}^{{\rm{X}},mbl}{\varpi ^m_{aa'q}}{e^{{\rm{j}}2\pi \left( {lT\frac{{f_{\rm c}^m}}{{{f_{\rm c}}}}{f_{aa'q}} - b{\Delta _f}{\tau _{aa'q}}} \right)}}}_{\mu _{aa'q}^{{\rm{X,}}mbl}[n]}}\nonumber\\
& +u_{aq,mbl}^{{\rm{X}},{\rm{R}}}.
 \label{eqappendix20}
 \end{align}

Then, we define $\mathbf{{y}}_{aq}^{{\rm{X}},{\rm{R}}}$ as the vector consisting of the elements ${y}_{aq,mbl}^{{\rm{X}},{\rm{R}}}$ for $0 \leq b \leq B_m^{\rm{X}} - 1$, $0 \leq l \leq L_m^{\rm{X}} - 1$, and $m\in{\mathbb S}_d^{\rm X}$.
Also, let $\mathbf{{y}}_q^{\rm{X,R}}$ denote the sensing-signal vector of target $q$, which is constructed by aggregating $\mathbf{{y}}_{aq}^{{\rm{X}},{\rm{R}}}$ over all $a \in {\mathbb A}_d^{\rm{X}}$.
Moreover, the motion parameters of target $q$ are represented by ${{\bm{\omega }}_q} =\left[ {{\left[ {{\tau _{aa'q}}} \right]}_{a,a' \in {{\mathbb A}_d^{\rm{X}}}}},{\left[ {{f_{aa'q}}} \right]}_{a,a' \in {{\mathbb A}_d^{\rm{X}}}} \right]$, and the unknown sensing parameters of target $q$ are represented by ${\bf x}_q^{\rm R}$ defined in \eqref{eq5}. The FIM with respect to ${\bf x}_q^{\rm R}$ is given by
 \begin{align}{{\bf{F}}_q^{\rm X}}[n]{\rm{ = }}\frac{{\partial {{\bm{\omega }}_q}}}{{\partial {{\bf x}_q^{\rm R}}}}\left[\!\! {\begin{array}{*{20}{c}}
{{{\bf{F}}_q^{{\rm X},\tau \tau }}}&\!\!\!{{{\bf{F}}_{q}^{{\rm X},\tau f}}}\\
\left({{{\bf{F}}_q^{{\rm X},\tau f}}}\right)^{\mathsf T}&\!\!\!{{{\bf{F}}_q^{{\rm X},ff}}}
\end{array}} \right]{\left( {\frac{{\partial {{\bm{\omega }}_q}}}{{\partial {{\bf x}_q^{\rm R}}}}} \right)^{\rm{T}}}\!\!\!\in{\mathbb R}^{4\times4},\label{eq11}
 \end{align}
 where
$\frac{{\partial {{\bm{\omega }}_q}}}{{\partial {{\bf x} _q^{\rm{R}}}}} $
  and the $(\ell,j)$-th elements of $ {\bf{F}}_q^{\tau \tau }$,  $ {\bf{F}}_q^{\tau f }$, and $ {\bf{F}}_q^{ff }$ are given by
 \begin{align}
 &\frac{{\partial {{\bm{\omega }}_q}}}{{\partial {\bf{x}}_q^{\rm{R}}}}  = \left[ {\begin{array}{*{20}{c}}
{{{\left[ {\frac{{\partial {\tau _{aa'q}}}}{{\partial x_q^{{\rm{R,x}}}}}} \right]}_{a,a' \in {{\mathbb A}_d^{\rm{X}}}}}}&{{{\left[ {\frac{{\partial {f_{aa'q}}}}{{\partial x_q^{{\rm{R,x}}}}}} \right]}_{a,a' \in {{\mathbb A}_d^{\rm{X}}}}}}\\
{{{\left[ {\frac{{\partial {\tau _{aa'q}}}}{{\partial x_q^{{\rm{R,y}}}}}} \right]}_{a,a' \in {{\mathbb A}_d^{\rm{X}}}}}}&{{{\left[ {\frac{{\partial {f_{aa'q}}}}{{\partial x_q^{{\rm{R,y}}}}}} \right]}_{a,a' \in {{\mathbb A}_d^{\rm{X}}}}}}\\
0&{{{\left[ {\frac{{\partial {f_{aa'q}}}}{{\partial v_q^{{\rm{R,x}}}}}} \right]}_{a,a' \in {{\mathbb A}_d^{\rm{X}}}}}}\\
0&{{{\left[ {\frac{{\partial {f_{aa'q}}}}{{\partial v_q^{{\rm{R,y}}}}}} \right]}_{a,a' \in {{\mathbb A}_d^{\rm{X}}}}}}
\end{array}} \right]
\in \mathbb R^{4\times 2\tilde A^2}, \nonumber\\
& {\left[{\bf F}_q^{{\rm X},\zeta\xi}\right]_{\ell j}}
=
{\rm E}
\left[
-
\frac{\partial^2\ln{\cal P}
\left({\bf y}_q^{{\rm X},{\rm R}}\mid{\bm\omega}_q\right)}
{\partial \zeta_{aa'q}\partial \xi_{\tilde a\tilde a'q}}
\right],
  \end{align}
respectively, for  $\zeta,\xi \in \left\{ {\tau ,f} \right\}$,  
where $\ell=(\iota(a)-1)\tilde A+\iota(a')$ and
$j=(\iota(\tilde a)-1)\tilde A+\iota(\tilde a')$, with
$\iota(\cdot)$ denoting the local AP index in ${\mathbb A}_d^{\rm X}$ and $\tilde A=\left|{\mathbb A}_d^{\rm X} \right|$.
Besides, we have 
\begin{align}
&\ln {\cal P}\left( {{\bf{  y}}_q^{{\rm{X}},{\rm{R}}}\left| {{\bm{\omega}}_q} \right.} \right)  \nonumber \\
\propto&-\! \!  \sum\limits_{a \in {{\mathbb A}_d^{\rm{X}}}} \!  {\sum\limits_{m \in {{\mathbb S}_d^{\rm{X}}}}\!   {\sum\limits_{b = 0}^{B_m^{\rm{X}} - 1} \!  {\sum\limits_{l = 0}^{L_m^{\rm{X}} - 1} {\frac{1}{\sigma_a^{\rm A} }{{\left| {y_{aq,mbl}^{{\rm{X}},{\rm{R}}}\! -\!\!  \sum\limits_{a' \in {{\mathbb A}_d^{\rm{X}}}} {\mu _{aa'q}^{{\rm{X,}}mbl}} } \right|}^2}} } } }.
 \end{align}

\subsubsection{FIM Calculation}
The FIM in \eqref{eq17} is derived based on \eqref{eq11}.
First, the first derivative of $\ln {\cal P}\left( {{\bf{y}}_q^{{\rm{X}},\rm{R}}\left| {{\bm{\omega}}_q} \right.} \right)$ with respect to ${\tau _{aa'q}}$ is
\begin{align}
&\frac{{\partial \ln {\cal P}\left( {{\bf{y}}_q^{{\rm{X}},{\rm{R}}}\left| {{{\bm{\omega }}_q}} \right.} \right)}}{{\partial {\tau _{aa'q}}}} = \frac{1}{{\sigma _a^{\rm{A}}}}\sum\limits_{m,b,l}^{} {\left( {\left( {y_{aq,mbl}^{{\rm{X}},{\rm{R}}} - \sum\limits_{a'} {\mu _{aa'q}^{{\rm{X}},mbl}} } \right)} \right.} \nonumber \\
& \times \! \!\frac{{\partial {{\left( {\mu _{aa'q}^{{\rm{X}},mbl}} \right)}^*}}}{{\partial {\tau _{aa'q}}}}\!\! +\!\! \left. {{{\left( {y_{aq,mbl}^{{\rm{X}},{\rm{R}}}\! - \!\sum\limits_{a'} {\mu _{aa'q}^{{\rm{X}},mbl}} } \right)\!}^*}\!\frac{{\partial \mu _{aa'q}^{{\rm{X}},mbl}}}{{\partial {\tau _{aa'q}}}}} \right).
\end{align}

Then, the $(\ell,j)$-th element of ${\bf F}_q^{{\rm X},\tau\tau}$ can be rewritten as
\begin{align}
&{\rm E}\left[
- \frac{\partial^2 \ln {\cal P}
\left(
{\bf y}_q^{{\rm X},{\rm R}}
\mid
{\bm \omega}_q
\right)}
{\partial \tau_{aa'q}\partial \tau_{\tilde a\tilde a'q}}
\right]
\nonumber\\
=&
\begin{cases}
\displaystyle
\frac{2}{\sigma_a^{\rm A}}
{\rm E}\left[
\sum_{m,b,l}
{\rm Re}\left(
\frac{\partial \mu_{aa'q}^{{\rm X},mbl}}
{\partial \tau_{aa'q}}
\frac{\partial
\left(\mu_{\tilde a\tilde a'q}^{{\rm X},mbl}\right)^*}
{\partial \tau_{\tilde a\tilde a'q}}
\right)
\right],
& a=\tilde a,\\[2mm]
0,
& a\ne \tilde a,
\end{cases}
\nonumber\\
=&
\begin{cases}
\displaystyle
\frac{2}{\sigma_a^{\rm A}}
{\rm E}\!\left[
\sum_{m,b,l}
{\rm Re}\!\left(
\begin{aligned}
&\tilde s_{a'q}^{{\rm X},mbl}
\left(\tilde s_{\tilde a'q}^{{\rm X},mbl}\right)^*
\\
&\times
\varpi_{aa'q}^{m}
\left(\varpi_{a\tilde a'q}^{m}\right)^*
\\
&\times
e^{{\rm j}2\pi\Delta_{\ell j}^{mbl}}(2\pi b\Delta_f)^2
\end{aligned}
\right)
\right],
& a=\tilde a,\\[4mm]
0,
& a\ne \tilde a,
\end{cases}
\nonumber\\
\approx&
\begin{cases}
\displaystyle
\frac{2}{\sigma_a^{\rm A}}
{\rm E}\left[
\sum_{m,b,l}
(2\pi b\Delta_f)^2
\left|
\varpi_{aa'q}^{m}
\tilde s_{a'q}^{{\rm X},mbl}
\right|^2
\right],
& \ell=j,\\[2mm]
0,
& \ell\ne j,
\end{cases}
\nonumber\\
=&
\begin{cases}
\displaystyle
\frac{
\sum_{m,b,l}
8(\pi b\Delta_f)^2
\lambda_{aa'q}^{\rm R}
\Gamma_{a'qm}^{{\rm X},{\rm R}}
}
{\sigma_a^{\rm A}},
& \ell=j,\\[2mm]
0,
& \ell\ne j,
\end{cases}
\nonumber\\
=&
\begin{cases}
\displaystyle
\frac{1}{\sigma_a^{\rm A}}
\bar W_{aa'q}^{\tau\tau},
& \ell=j,\\[2mm]
0,
& \ell\ne j.
\end{cases}
\label{eq51}
\end{align} 
where  $
\Delta_{\ell j}^{mbl}
=
\frac{f_{\rm c}^m}{f_{\rm c}}
\left(
f_{aa'q}
-
f_{\tilde a\tilde a'q}
\right)lT
-
b\Delta_f
\left(
\tau_{aa'q}
-
\tau_{\tilde a\tilde a'q}
\right)$ and $\bar W_{aa'q}^{\tau \tau }$ is defined in \eqref{eqWtt}.
Similarly, we have
 \begin{align}
&\!\!{\rm{E}}\left( { - \frac{{{\partial ^2}\ln {\cal P}\left( {{\bf{  y}}_q^{{\rm X},\rm{R}}\left| {{{\bm{\omega }}_q}} \right.} \right)}}{{\partial {f_{aa'q}}\partial {f_{\tilde a\tilde a'q}}}}} \right)
\approx \left\{ {\begin{array}{*{20}{l}}
{\frac{1}{\sigma_a^{\rm A}}{{\bar W_{aa'q}^{ff}}},} &\ell =  j,\\
{0,}&\ell \ne   j,
\end{array}} \right.\\
&\!\!{\rm{E}}\left( { - \frac{{{\partial ^2}\ln {\cal P}\left( {{\bf{y}}_q^{{\rm X},\rm{R}}\left| {{{\bm{\omega }}_q}} \right.} \right)}}{{\partial {\tau _{aa'q}}\partial {f_{\tilde a\tilde a'q}}}}} \right) 
 \approx \left\{ {\begin{array}{*{20}{l}}
{\frac{-1}{\sigma_a^{\rm A}}{{  \bar W_{aa'q}^{\tau f} }},}&\ell =  j,\\
{0,}&\ell \ne  j,
\end{array}} \right.
\end{align}
where $\bar W_{aa'q}^{ff}$ and $\bar W_{aa'q}^{\tau f}$    are defined in \eqref{eqWff} and \eqref{eqWtf}, respectively.
Finally, we obtain the FIM in \eqref{eq17}.

\ifCLASSOPTIONcaptionsoff

\fi
  \bibliography{SPT}
\bibliographystyle{IEEEtran}

\begin{IEEEbiography}[{\includegraphics[width=1in,height=1.25in, clip,keepaspectratio]{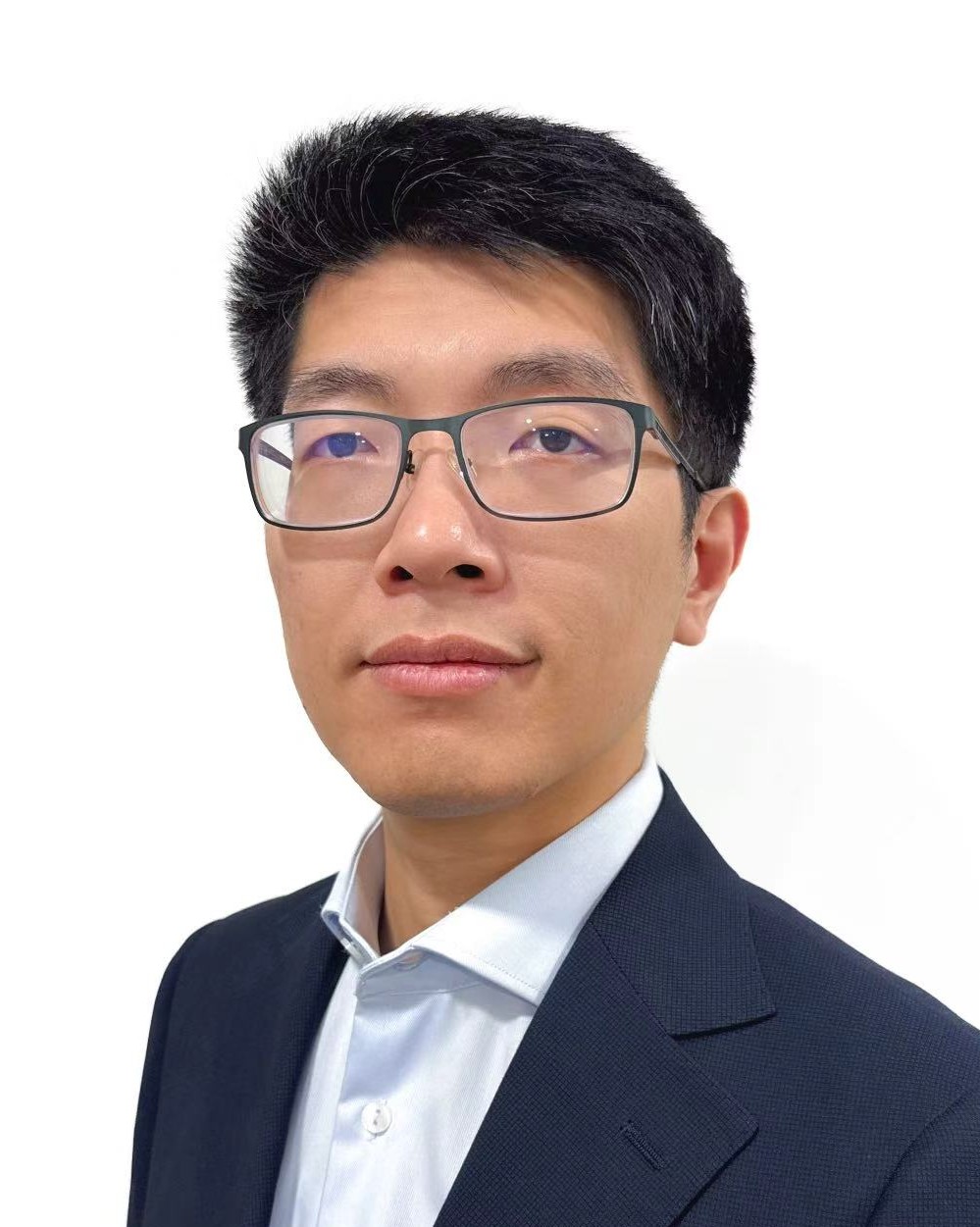}}]{Jie Chen} (Member, IEEE)  received his Ph.D. degree in communication and information systems from the University of Electronic Science and Technology of China (UESTC), Chengdu, China, in 2021.
From 2019 to 2020, he was a Visiting Ph.D. Student at the University of Toronto, Toronto, ON, Canada. 
From 2022 to 2026, he was a Post-Doctoral Associate with the Department of Electrical and Computer Engineering, Western University, London, ON, Canada. 
He is currently a Research Scientist with the Department of Electrical and Computer Engineering, University of New Brunswick, Fredericton, NB, Canada.
His research interests include integrated sensing and communications, transceiver design for Internet-of-Things, and machine learning for wireless communications.
He was the recipient of the Journal of Communications and Information Networks (JCIN) Best Paper Award in 2021.
 
\end{IEEEbiography}

\begin{IEEEbiography}[{\includegraphics[width=1in,height=1.25in, clip,keepaspectratio]{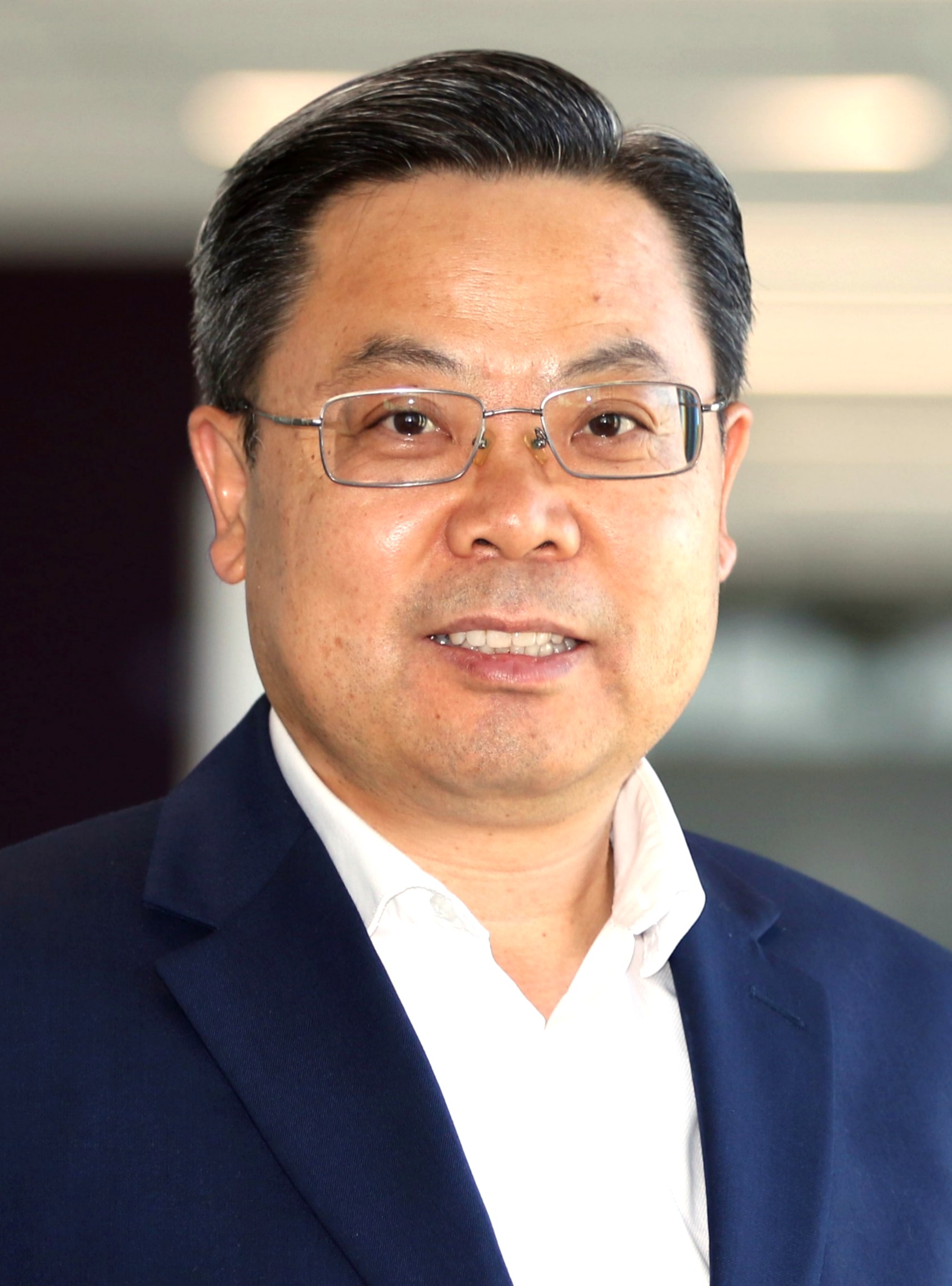}}]{Xianbin Wang} (Fellow, IEEE) received his Ph.D. degree in electrical and computer engineering from the National University of Singapore in 2001.

He has been with Western University, Canada, since 2008, where he is currently a Distinguished University Professor and Tier-1 Canada Research Chair in Trusted Communications and Computing.
Prior to joining Western University, he was with the Communications Research Centre Canada as a Research Scientist and later a Senior Research Scientist from 2002 to 2007. From 2001 to 2002, he was a System Designer at STMicroelectronics. His current research interests include 5G/6G technologies, Internet of Things, machine learning, communications security, digital twin, and intelligent communications. He has over 700 highly cited journals and conference papers, in addition to over 30 granted and pending patents and several standard contributions.

Dr. Wang is a Fellow of the Canadian Academy of Engineering and a Fellow of the Engineering Institute of Canada. He has received many prestigious awards and recognitions, including the IEEE Canada R. A. Fessenden Award, Canada Research Chair, Engineering Research Excellence Award at Western University, Canadian Federal Government Public Service Award, Ontario Early Researcher Award, and twelve Best Paper Awards. He is currently a member of the Senate, Senate Committee on Academic Policy and Senate Committee on University Planning at Western. He also serves on NSERC Discovery Grant Review Panel for Computer Science. He has been involved in many flagship conferences, including IEEE GLOBECOM, ICC, VTC, PIMRC, WCNC, CCECE, and ICNC, in different roles, such as General Chair, TPC Chair, Symposium Chair, Tutorial Instructor, Track Chair, Session Chair, and Keynote Speaker. He was nominated as an IEEE Distinguished Lecturer multiple times by different societies including BTS, ComSoc and VTS. He serves/has served as the Editor-in-Chief, Associate Editor-in-Chief, Area Editor, and editor/associate editor for over ten journals. He was the Chair of the IEEE ComSoc Signal Processing and Computing for Communications (SPCC) Technical Committee and the Central Area Chair of IEEE Canada.
\end{IEEEbiography}

\end{document}